\documentclass[sigconf]{acmart}
\usepackage{booktabs}    
\usepackage{multirow}    
\usepackage{makecell}    
\usepackage{adjustbox}   
\usepackage{subcaption} 

\AtBeginDocument{%
  }
\usepackage{enumitem}
\setlist[itemize]{leftmargin=12pt}

\usepackage{xcolor}

\copyrightyear{2025}
\acmYear{2025}
\setcopyright{acmlicensed}\acmConference[MM '25]{Proceedings of the 33rd ACM International Conference on Multimedia}{October 27--31, 2025}{Dublin, Ireland}
\acmBooktitle{Proceedings of the 33rd ACM International Conference on Multimedia (MM '25), October 27--31, 2025, Dublin, Ireland}
\acmDOI{10.1145/3746027.3755597}
\acmISBN{979-8-4007-2035-2/2025/10}

\usepackage{multirow}
\usepackage{adjustbox}
\begin{document}

\title{Dual-Phase Playtime-guided Recommendation: Interest Intensity Exploration and Multimodal Random Walks}

\author{Jingmao Zhang}
\authornote{Both authors contributed equally to this research.}
\affiliation{%
  \institution{Harbin Institute of Technology}
  \city{Shenzhen}
  \country{China}}
\email{220810426@stu.hit.edu.cn}

\author{Zhiting Zhao}
\authornotemark[1]
\affiliation{%
  \institution{Harbin Institute of Technology}
  \city{Shenzhen}
  \country{China}}
\email{24s151185@stu.hit.edu.cn}

\author{Yunqi Lin}
\affiliation{%
  \institution{Harbin Institute of Technology}
  \city{Shenzhen}
  \country{China}}
\email{210110519@stu.hit.edu.cn}
 
\author{Jianghong Ma}
\authornote{Corresponding author}
\affiliation{%
  \institution{Harbin Institute of Technology}
  \city{Shenzhen}
  \country{China}}
\email{majianghong@hit.edu.cn}

\author{Tianjun Wei}
\authornotemark[2]
\affiliation{%
  \institution{Nanyang Technological University}
  \country{Singapore}}
\email{tjwei2-c@my.cityu.edu.hk}

\author{Haijun Zhang}
\affiliation{%
  \institution{Harbin Institute of Technology}
  \city{Shenzhen}
  \country{China}}
\email{hjzhang@hit.edu.cn}

\author{Xiaofeng Zhang}
\affiliation{%
  \institution{Harbin Institute of Technology}
  \city{Shenzhen}
  \country{China}}
\email{zhangxiaofeng@hit.edu.cn}

\renewcommand{\shortauthors}{}

\begin{abstract}
The explosive growth of the video game industry has created an urgent need for recommendation systems that can scale with expanding catalogs and maintain user engagement. While prior work has explored accuracy and diversity in recommendations, existing models underutilize playtime, a rich behavioral signal unique to gaming platforms, and overlook the potential of multimodal information to enhance diversity. In this paper, we propose $DP^2Rec$, a novel Dual-Phase Playtime-guided Recommendation model designed to jointly optimize accuracy and diversity. First, we introduce a playtime-guided interest intensity exploration module that separates strong and weak preferences via dual-beta modeling, enabling fine-grained user profiling and more accurate recommendations. Second, we present a playtime-guided multimodal random walks module that simulates player exploration using transitions guided by both playtime-derived interest similarity and multimodal semantic similarity. This mechanism preserves core preferences while promoting cross-category discovery through latent semantic associations and adaptive category balancing. Extensive experiments on a real-world game dataset show that $DP^2Rec$ outperforms existing methods in both recommendation accuracy and diversity. The dataset and source code are released at {https://github.com/zqxwcevrtyui/DP2Rec}
\end{abstract}



\maketitle

\section{Introduction}
The video game industry is growing rapidly, with 18,949 new games released on Steam in 2024—a 32.42\% increase from 2023. Steam also hit a record 39.3 million concurrent users in December 2024, reflecting the rising demand driven by rich visual and textual content.
This surge underscores the need for domain-specific recommendation systems~\cite{bharathipriya2021online,caroux2023presence,cheuque2019recommender,ikram2022multimedia,perez2020hybrid,yang2022large} to help users navigate expanding game catalogs and enhance engagement.
This accelerated market expansion underscores the critical need for advanced, domain-specific recommendation systems in the gaming domain~\cite{bharathipriya2021online,caroux2023presence,cheuque2019recommender,ikram2022multimedia,perez2020hybrid,yang2022large}. Such systems are instrumental in helping users navigate the growing volume of releases by recommending preferred games, thereby enhancing user experience and platform retention.

While accuracy, driven by users’ historical behavior \cite{xu2024learning,chen2024post,lv2025dynamic,zhao2024denoising,sun2024self}, remains the primary focus of recommender systems, recent work has broadened the evaluation scope to include diversity, which is essential for reducing redundancy and promoting discovery \cite{zheng2024diversity,chen2025dlcrec,xiao2024divnet,yin2024simple,zhao2024can}. Striking a balance between the two ensures sustained engagement without sacrificing personalization \cite{coppolillo2024relevance,peng2024reconciling,eskandanian2020using,duricic2023beyond}. While these dimensions have been extensively studied in general recommendation settings, domain-specific adaptations—such as game recommendation—have seen incremental advances. For instance, SCGRec \cite{SCGRec} boosts accuracy via social-contextualized graph neural networks; DRGame \cite{liu2024drgame} augments diversity through clustered node selection; and CPGRec \cite{li2024category} imposes category-aware constraints to harmonize both objectives.

Crucially, \textit{\textbf{playtime}}—a behavioral signal unique to gaming platforms like Steam—provides an untapped granularity for preference modeling. Unlike discrete interactions (e.g., clicks or ratings) in traditional domains, playtime offers a temporal proxy for latent interest, yet current methods fail to exploit its full potential. CPGRec disregards playtime entirely, relying only on categorical graph construction, while SCGRec and DRGame inadequately model interaction strength, failing to resolve fine-grained preference distinctions encoded in playtime patterns. Specifically, two key challenges persist in game recommendation research:

\begin{itemize}
\item \textbf{Playtime-aware Dual-Interest Modeling.} Current game recommenders underutilize playtime data, often treating interactions as binary preferences. While some methods incorporate playtime, they miss the interest intensity spectrum it encodes - prolonged engagement indicates strong interest while brief interactions suggest weak interest. Our analysis (Section~\ref{sec:User Interest Intensity Modeling}) shows two distinct playtime-driven interest types, necessitating separation of core preferences from casual ones. Though traditional domains {\cite{bai2025chime,wei2023multi,liu2023modeling,cho2023dynamic}} model dynamic interests, they lack playtime integration for gaming contexts, highlighting the need for fine-grained interest differentiation.

\item \textbf{Cross-Category Diversity Through Multimodal Similarity.} While multimodal learning is prevalent in graph \cite{wen2023graph,liu2023information} and recommenders, existing methods (both general and game-specific) primarily use modalities (text, images) for accuracy, overlooking diversity potential. Current approaches limit content similarity to historical preferences, causing homogenized recommendations. We show multimodal representations can reveal latent cross-category associations, enabling diverse yet relevant recommendations. Our analysis (Section~\ref{sec:modal-cross-category}) demonstrates that high multimodal similarity often occurs across distinct categories, proving modalities can transcend categorical constraints.
\end{itemize}
To address the above challenges, we propose a novel model: Dual-Phase Playtime-guided Recommendation: Interest Intensity Exploration and Multimodal Random Walks ($DP^2Rec$)—designed to improve both the accuracy and diversity of game recommendations. \textbf{First}, we introduce a \textit{playtime-guided interest intensity exploration} module to enhance accuracy, which identifies strong and weak preferences in players’ gameplay duration via dual-beta distribution separation based on Expectation-Maximization (EM) algorithm \cite{dempster1977maximum}, thereby enabling fine-grained modeling of user interest levels. \textbf{Second}, to enhance diversity, we present a \textit{playtime-guided multimodal random walks} module. This module uses a novel random walks mechanism to simulate the user exploration process where node transitions are jointly determined
by semantic alignment (via multimodal embeddings) and behavioral alignment (via playtime-based interest). The dual-alignment preserves users' core interest while enabling cross-category discovery through multimodal space traversal and adaptive category balance coefficients. Collectively, these two modules lead to significant improvements in both recommendation accuracy and diversity.

The main contributions can be summarized as follows:

\begin{itemize}
\item To our knowledge, this work presents the \textbf{first} framework in game recommendation systems that leverages playtime dynamics to explicitly optimize both accuracy and diversity. 

\item Through in-depth data analysis, we uncover two key insights: (1) playtime reflects distinct levels of user interest intensity, supported by theoretical validation, revealing fine-grained preference structures; and (2) multimodal representations of games uncover latent semantic links across categories, enabling effective cross-category exploration. 

\item Building on the above-mentioned insights, we design two complementary modules: an interest intensity exploration module to boost accuracy, and a multimodal random walk module to enhance diversity. Theoretical analysis further substantiates the effectiveness of the proposed diversity-enhancing mechanism. 

\item Extensive experiments on real-world game data demonstrate that our method outperforms state-of-the-art recommendation techniques in both accuracy and diversity.
\end{itemize}

\section{Related Work}
\subsection{Game Recommendation}
Early game recommendation research focused on Collaborative Filtering (CF) and Content-Based Filtering (CBF). GAMBIT~\cite{anwar2017game} applies CF by modeling users and games separately, while BharathiPriya \textit{et al}. \cite{bharathipriya2021online} combine CF and CBF, incorporating playtime for implicit ranking. With the rise of deep learning, models like DeepNN and DeepFM \cite{cheuque2019recommender} have outperformed traditional methods on the Steam dataset. Graph Neural Networks (GNNs) have also gained traction for their strength in modeling complex relations. For instance, SCGRec~\cite{yang2022large} enhances accuracy using game context and social links, while CPGRec~\cite{li2024category} improves propagation by enforcing strict game connections.

\textbf{Discussion.} Our approach differs from existing methods in two key ways: (1) As far as we know, we are the \textbf{first} to model player behavior by extracting dual-interest signals from playtime; and (2) as far as we know, we are also the \textbf{first} to introduce a random walks-based item selection strategy for game recommendation. 

\subsection{Diversified Recommendation}
Since Ziegler et al.\ \cite{ziegler2005improving} introduced diversified recommendation, many follow-up approaches have been developed to enhance diversity. Re-ranking methods like Maximum Marginal Relevance (MMR) \cite{carbonell1998use, peska2022towards, abdool2020managing, lin2022feature} and Determinantal Point Process (DPP) \cite{chen2018fast, gan2020enhancing, huang2021sliding} enhance diversity by adjusting output rankings. More recently, GNN-based methods~\cite{tian2022reciperec} leverage high-order graph structures for diversity. DGCN~\cite{zheng2021dgcn} incorporates category-aware sampling, while DGRec~\cite{yang2023dgrec} introduces submodular neighbor selection. DRGame~\cite{liu2024drgame} explores cluster-based diversification, and DDGraph~\cite{ye2021dynamic} dynamically updates the user-item graph for diverse recommendations. Emerging LLM-based approaches, such as DLCRec~\cite{chen2024dlcrec}, offer controllable frameworks to balance category diversity using language models.

\textbf{Discussion.} To the best of our knowledge, we are the \textbf{first} to leverage multimodal similarity specifically to enhance recommendation diversity. Our playtime-guided multimodal random walks module integrates multimodal information and adopts a hierarchical exploration strategy. By modeling the dynamic shift of user interests through path-level cumulative similarity, it preserves personalization while effectively expanding cross-category discovery. 

\section{Data Analysis}
We begin with data analysis of the Steam gaming platform dataset\footnote{\url{https://store.steampowered.com}}. 
\subsection{User Interest Intensity Modeling Based on Dual-Beta Distribution}
\label{sec:User Interest Intensity Modeling}

\subsubsection{Rationale for Dual-Beta Modeling.}
\label{sec:dual-beta}
To characterize users' interest intensity in games, we categorize player interactions into strong and weak interest interactions, using Beta distribution for precise modeling. First, players' gameplay time is normalized to the interval $[0,1]$ using the percentile method:
\begin{equation}
\hat t^u_i = \frac{\text{rank}(t^u_{i})}{\text{total\_players}(i)}
\end{equation}
where $t^{u}_i$ represents user $u$'s gameplay time for game $i$, $\text{rank}(t^u_i)$ denotes the ascending rank position of this time among all users who played game $i$ (longer time corresponds to larger value), and $\text{total\_players}(i)$ represents the total number of users who played game $i$. This normalization method converts time to relative percentiles, perfectly matching the domain of Beta distribution and forming an ideal modeling foundation.

The Beta distribution is determined by shape parameters $\alpha$ and $\beta$, with probability density function:
\begin{equation}
f(x; \alpha, \beta) = \frac{x^{\alpha - 1}(1 - x)^{\beta - 1}}{B(\alpha, \beta)}, \quad x \in (0,1)
\label{eq:beta}
\end{equation}
where $B(\alpha, \beta)$ is the Beta function\footnote{The percentile-based normalized playtime $\hat{t}^u_i \in [0,1]$ may include boundary values 0 or 1. To ensure compatibility with the open interval $(0,1)$ required by the Beta distribution and to avoid numerical instability, we apply a small offset $\epsilon$ (e.g., $10^{-6}$) to shift extreme values into $(0, 1)$.}. Notably, \textbf{strong interest} often concentrates in high percentiles (close to 1), which can be represented by Beta distributions where $\boldsymbol{\alpha > \beta}$, such as $\text{Beta}(5,2)$. With this parameter configuration, the density function peaks near 1, precisely reflecting users' behavior of investing long gameplay time in preferred games. Conversely, \textbf{weak interest} can be represented by Beta distributions where $\boldsymbol{\alpha < \beta}$, such as $\text{Beta}(2,5)$, with distributions skewed toward lower percentiles, matching users' shallow interaction characteristics.

Based on above discussions, we establish separate Beta distributions for each user's strong and weak interests, collectively called the dual-beta model. The probability density function of the dual-beta distribution is:
$$
f(x) = \pi f_1(x; \alpha_1, \beta_1) + (1-\pi) f_2(x; \alpha_2, \beta_2)
$$
where $\pi$ represents the weight of the strong interest distribution, and $f_1$ and $f_2$ are the Beta distribution density functions for strong and weak interests, respectively.

\subsubsection{Validation of Dual-Beta Modeling via KS Test. } We select users with no fewer than 10 game interactions from the original pool of 60,742 players, resulting in a total of 53,025 valid users as our study sample. These users have relatively complete behavioral records, better reflecting their intrinsic interest structures. For each user, we use Maximum Likelihood Estimation (MLE) to separately fit the Beta distribution parameters for strong and weak interests.

\begin{table}[htbp]
\centering
\caption{KS Fitting Test Statistics}

\label{tab:ks_test}
\begin{tabular}{ccc}
\hline
Successful Fits & Average KS Statistic & Average $p$-value \\
\hline
53003 / 53025 (99.96\%) & 0.0696 & 0.9638 \\
\hline
\end{tabular}
\end{table}

To evaluate the fitting quality, we apply the Kolmogorov–Smirnov (KS) test \cite{massey1951kolmogorov}. A fitting is considered successful if the p-value exceeds the significance level of 0.05. The statistical results are shown in Table \ref{tab:ks_test}. The experimental results indicate that among the 53,025 users, \textbf{53,003 users (99.96\%)} had their interest distributions accurately modeled by the dual-beta distribution. Notably, the \textbf{average p-value reached 0.9638, far exceeding the 0.05 threshold}, which strongly supports the effectiveness and accuracy of using dual-beta distributions for modeling user interests.

\begin{table}[htbp]
\centering
\caption{Beta Distribution Parameter Statistics} 
\label{tab:beta_params} 
\begin{tabular}{ccc}
\hline
Interest Type & Average $\alpha$ & Average $\beta$ \\
\hline
Strong Interest & 9.58 & 2.26 \\
Weak Interest & 4.68 & 8.37 \\
\hline
\end{tabular}
\end{table}

Further analysis of the fitted parameters yields the statistical characteristics for strong and weak interest distributions as shown in Table \ref{tab:beta_params}. From the parameter perspective, \textbf{strong interest} distributions generally exhibit a \textbf{left-skewed} shape with $\alpha > \beta$ (averaging $\text{Beta}(9.58, 2.26)$), which aligns with theoretical expectations and indicates players tend to engage deeply and consistently with games of strong interest. In contrast, \textbf{weak interest} distributions exhibit a \textbf{right-skewed} form with $\alpha < \beta$ (averaging $\text{Beta}(4.68, 8.37)$), reflecting shallow or occasional interactions. This also captures a subset of users with more exploratory behavioral patterns.

\subsection{Cross-Category Structure Analysis Based on Multimodal Game Representations}
\label{sec:modal-cross-category}
\begin{figure}[htbp]
    \hspace{-6mm}
    \begin{subfigure}{0.52\linewidth}
        \centering
        \includegraphics[width=\linewidth]{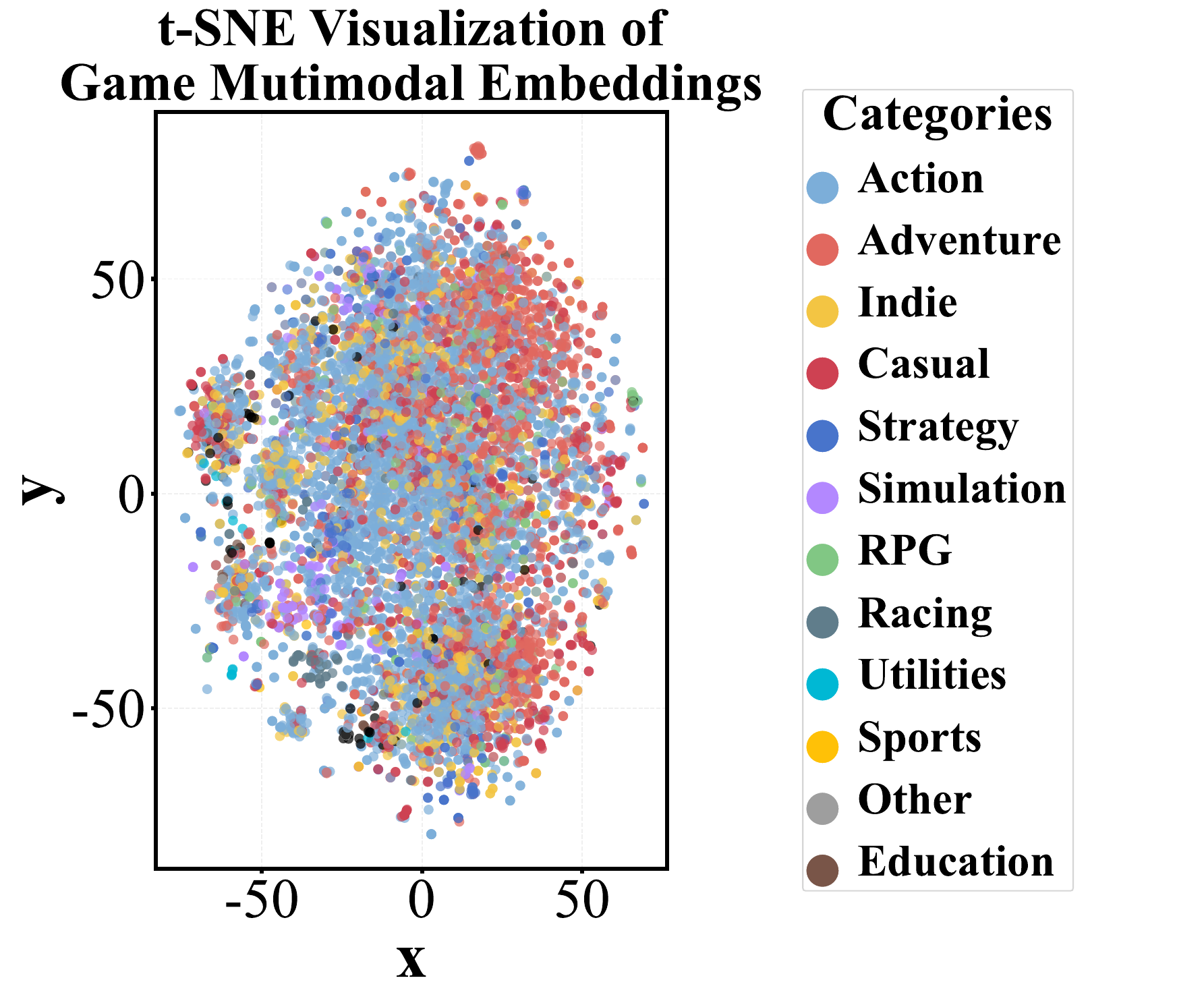}
        \caption{t-SNE visualization of game\\ multimodal embeddings}
        \label{fig:tsne}
    \end{subfigure}
    \hspace{-5mm}
    \begin{subfigure}{0.52\linewidth}
        \centering
        \includegraphics[width=\linewidth]{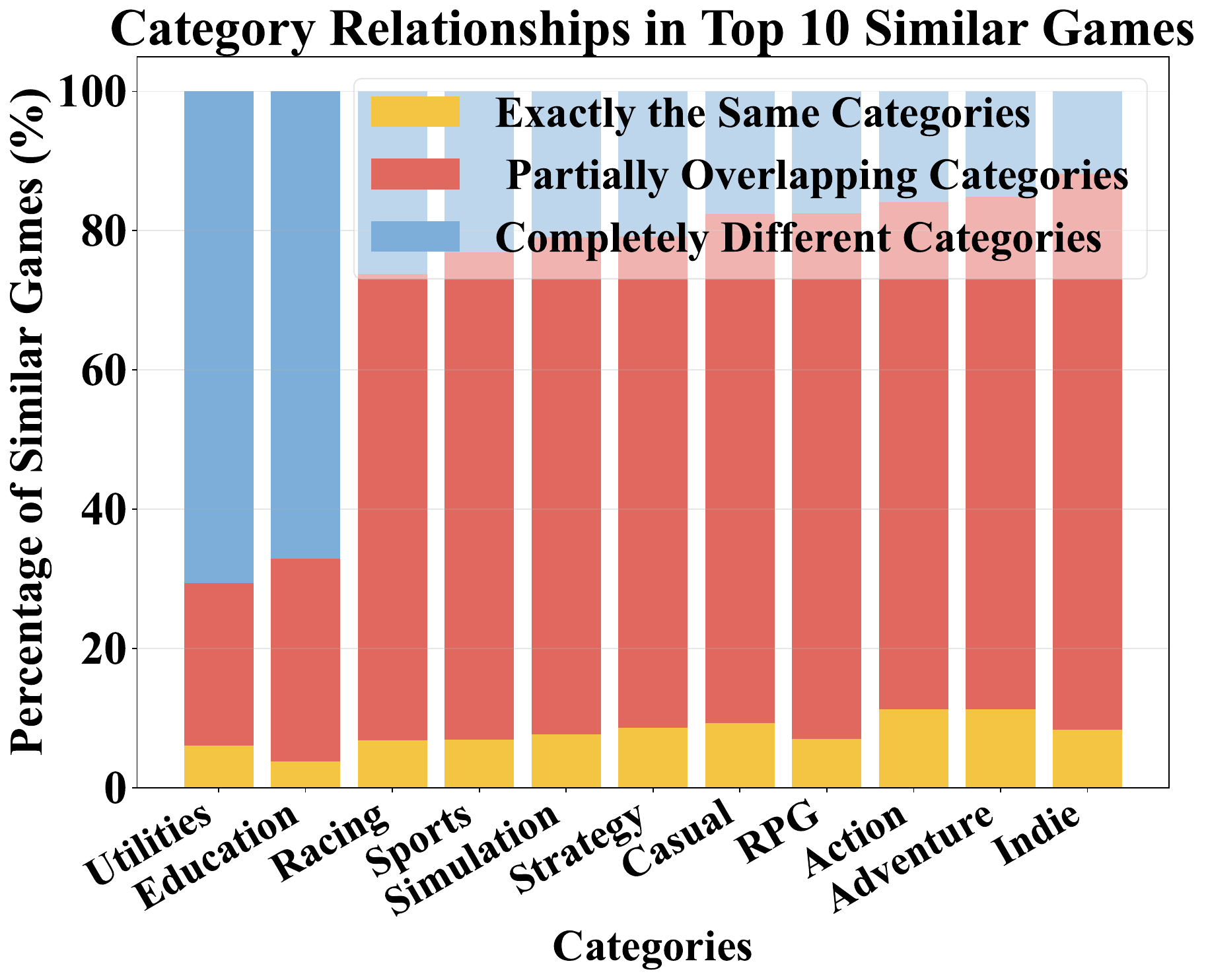}
        \caption{Category relationships in Top-10 modally similar games}
        \label{fig:distribution}
    \end{subfigure}
    \caption{Multimodal distribution and structural relations across game categories.}
    \label{fig:modal-category-analysis}
\end{figure}

To assess whether multimodal information facilitates cross-category exploration, we analyze the multimodal embeddings of games. We first use a large language model (LLM) to generate textual descriptions of game cover images and fuse them with the original game introductions. The fused text is then encoded using BERT \cite{devlin2019bert} to obtain a semantic embedding for each game. In this study, we adopt Qwen-2.5-VL-72B-Instruct \cite{bai2025qwen2}  as the LLM for both image captioning and text fusion, due to its strong contextual understanding and natural language generation capabilities, making it well-suited for aligning visual and textual modalities.

As shown in Figure~\ref{fig:tsne}, the t-SNE visualization of multimodal embeddings reveals that games from different categories do not form clearly separated clusters in the embedding space, but rather exhibit significant overlap. Figure~\ref{fig:distribution} further illustrates the category distribution of the top-10 games retrieved based on multimodal similarity. Specifically, for each category, we retrieve the top-10 similar games for every game in that category, then compute the average distribution of category relationships among those similar games. For example, in the \textit{Action} category, only 11.30\% of the similar games belong to exactly the same category, 72.82\% partially overlap, and 15.88\% fall into completely different categories. In contrast, in the \textit{Indie} category, 79.89\% of the similar games exhibit partial overlap. These results suggest that \textbf{multimodal similarity can support cross-category discovery while preserving semantic relevance}.

\section{Method}

\subsection{Overview}
\begin{figure*}
    \centering
    \includegraphics[width=1\linewidth]{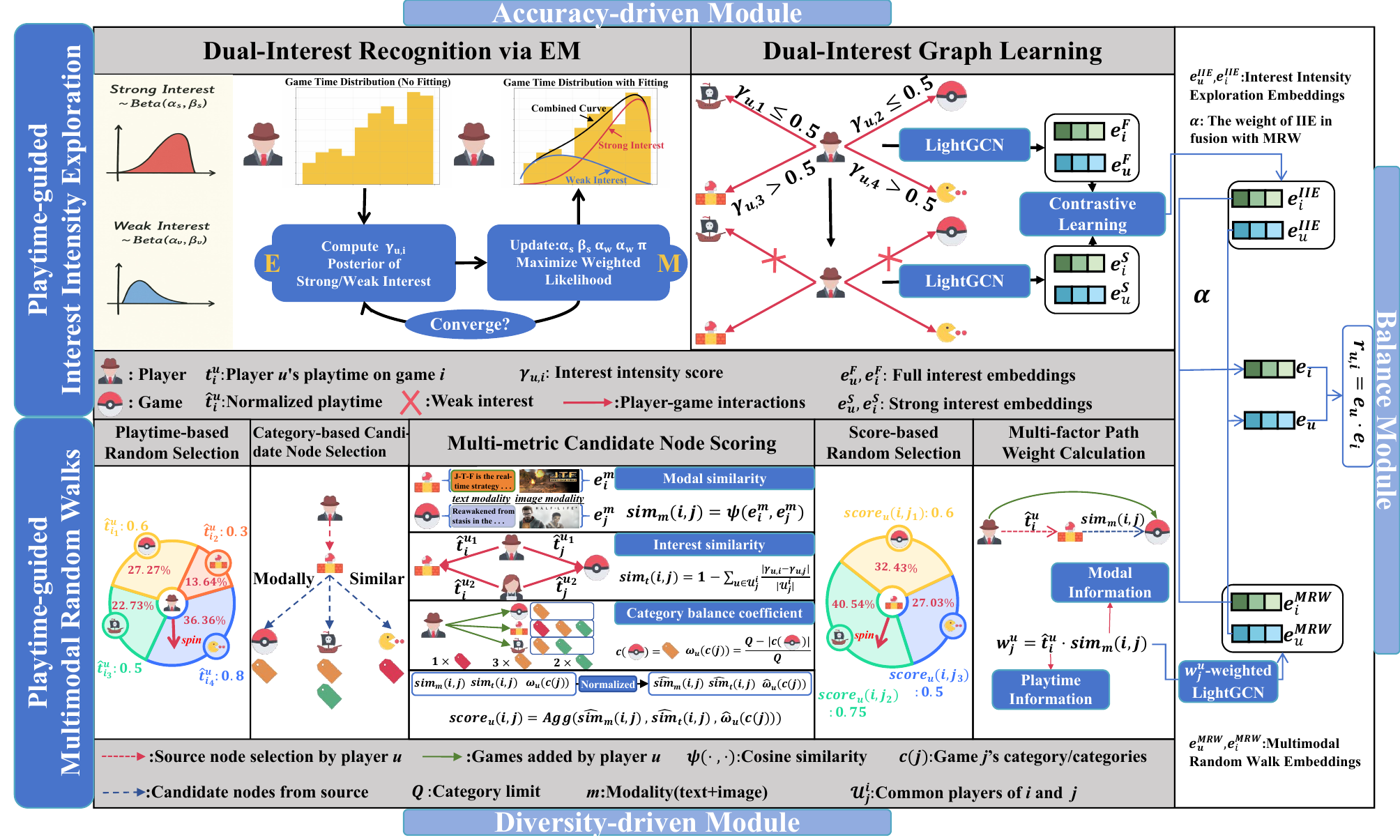}
    \caption{An illustration of the proposed $DP^2Rec$ model. The upper module applies \textit{Playtime-guided IIE}, where strong and weak interests are disentangled and encoded through two submodules. The lower module features a \textit{Playtime-guided MRW}, where multimodal and interest-based similarities guide user exploration through five submodules, preserving core preferences while enhancing diversity via cross-category discovery based on multimodal similarity and adaptive category balance coefficients. The balance module on the right fuses both modules using weighted integration to improve both accuracy and diversity.}
    \label{fig:model}    
\end{figure*}

As depicted in Figure \ref{fig:model}, the proposed $DP^2Rec$ model consists of two main modules: \textit{Playtime-guided Interest Intensity Exploration (IIE)} and \textit{Playtime-guided Multimodal Random Walks (MRW)}. The former is designed to enhance accuracy, whereas the latter promotes diversity. Lastly, a balance module is incorporated to fuse the result embeddings from the aforementioned modules.

\subsection{Playtime-guided Interest Intensity Exploration}

As detailed in Section \ref{sec:User Interest Intensity Modeling}, we model strong and weak interests using a dual-beta distribution tailored to the playtime signal. To disentangle these distinct interest types, we employ the EM algorithm, enabling a principled separation of underlying interest patterns. 

\subsubsection{Dual-Interest Recognition via EM}
The EM algorithm offers an efficient framework for parameter estimation in probabilistic models with latent variables from mixture distributions. By alternating between E and M steps, it uncovers hidden patterns from unlabeled data. In our case, EM is used to classify playtime distributions into strong and weak interests.

To model the distributional patterns of user playtime, we adopt a dual-beta mixture model within the EM algorithm:
\begin{equation}
p(\hat{t}^u_i) = \pi \cdot \text{Beta}(\hat{t}^u_i; \alpha_S, \beta_S) + (1 - \pi) \cdot \text{Beta}(\hat{t}^u_i; \alpha_W, \beta_W)
\end{equation}
where $p(\hat{t}^u_i)$ denotes the probability density function of player $u$’s normalized playtime for game $i$, and $\pi$ is the mixing weight for the strong interest component. The Beta parameters $(\alpha_S, \beta_S)$ and $(\alpha_W, \beta_W)$ correspond to strong and weak interests. 

We independently perform the EM algorithm on each player’s normalized playtime set $\{\hat{t}^u_i\}_{i \in \mathcal{I}_u}$, where $\mathcal{I}_u$ is the set of games interacted by player $u$. Initialization begins by sorting playtime in descending order and randomly selecting the top 40\% as the initial strong interest set, from which initial parameter estimates are derived. Interactions with zero playtime are treated as weak interest and excluded from EM updates. During the E-step, for each non-zero interaction $(u,i)$, the posterior probability of representing strong interest is calculated as: 
\begin{equation}
\gamma_{u,i}^{(S)} = \frac{\pi \cdot \text{Beta}(\hat{t}^u_i; \alpha_S, \beta_S)}{\pi \cdot \text{Beta}(\hat{t}^u_i; \alpha_S, \beta_S) + (1 - \pi) \cdot \text{Beta}(\hat{t}^u_i; \alpha_W, \beta_W)}
\end{equation}
where the Beta probability density function is defined in Equation \ref{eq:beta}.
In the M-step, we treat the posterior probabilities from the E-step as soft labels and perform weighted MLE to update the distribution parameters:
\begin{equation}
\alpha_S = \frac{\sum_{i \in \mathcal{I}_u} \gamma_{u,i}^{(S)} \ln\hat{t}^u_i}{-\sum_{i \in \mathcal{I}_u} \gamma_{u,i}^{(S)} \ln\hat{t}^u_i + \sum_{i \in \mathcal{I}_u} \gamma_{u,i}^{(S)} \ln(1-\hat{t}^u_i)}
\end{equation}
\begin{equation}
\beta_S = \alpha_S \cdot \frac{\sum_{i \in \mathcal{I}_u} \gamma_{u,i}^{(S)} \ln(1-\hat{t}^u_i)}{\sum_{i \in \mathcal{I}_u} \gamma_{u,i}^{(S)} \ln\hat{t}^u_i}
\end{equation}

The mixing weight $\pi$ is updated as the average of the posterior probabilities:
\begin{equation}
\pi = \frac{1}{|\mathcal{I}_u|}\sum_{i \in \mathcal{I}_u} \gamma_{u,i}^{(S)}
\end{equation}

By iteratively performing E and M steps until convergence, we obtain the final posterior probability $\gamma_{u,i}^{(S)}$ for each interaction $(u,i)$ being classified as strong interest.

\subsubsection{Dual-Interest Graph Learning}

Based on the posterior probabilities, interactions with $\gamma_{u,i}^{(S)}>0.5$ are classified as strong interest; others are considered weak. We construct two complementary graph views: the full interest graph $\mathcal{G}_{F} = (\mathcal{V}, \mathcal{E}_F)$, which includes all  interactions, and the strong interest graph $\mathcal{G}_{S} = (\mathcal{V}, \mathcal{E}_S)$, which retains only edges satisfying $\gamma_{u,i}^{(S)}>0.5$. Notably, (1) $\boldsymbol{\mathcal{G}_{S}}$ filters out weak interactions to highlight \textbf{users’ core interests}, while (2) the $\boldsymbol{\mathcal{G}_F}$ retains both dominant and exploratory signals, offering a \textbf{comprehensive view of user behavior}.

To enhance strong-interest signals while effectively leveraging the supplementary information in weak-interest interactions, we apply contrastive learning between the embeddings derived from both graph views, formulated by the following loss function:

\begin{equation}
\begin{aligned}
\mathcal{L}_{ssl,user}&=\sum_{u\in\mathcal{U}}-\log\frac{\exp(\phi\left(\mathbf{e}^F_{u},\mathbf{e}_{u}^{S}\right)/\tau)}{\sum_{v\in\mathcal{U}}\exp(\phi(\mathbf{e}^F_{u},\mathbf{e}_{v}^{S})/\tau)}\\
\mathcal{L}_{ssl,item}&=\sum_{i\in \mathcal{I}}-\log\frac{\exp(\phi\left(\mathbf{e}^F_{i},\mathbf{e}_{i}^{S}\right)/\tau)}{\sum_{j\in \mathcal{I}}\exp(\phi\left(\mathbf{e}^F_{i},\mathbf{e}_{j}^{S}\right)/\tau)}
\end{aligned}
\end{equation}
where $\mathbf{e}^F_{u}$ and $\mathbf{e}^F_{i}$ are the embeddings learned from the full graph $\mathcal{G}_F$, and $\mathbf{e}_{u}^{S}$ and $\mathbf{e}_{i}^{S}$ are the corresponding embeddings from the strong interest graph $\mathcal{G}_S$. $\phi(\cdot,\cdot)$ denotes the cosine similarity , and $\tau$ is the temperature hyperparameter.

After optimization, we obtain enhanced representations $\mathbf{e}_{u}^{IIE}$ and $\mathbf{e}_{i}^{IIE}$ that fuse dual-interest signals. These embeddings strengthen users’ core interests while retaining potentially valuable signals from weak interactions, providing richer and more personalized representations for downstream recommendation tasks.

The overall contrastive loss to balance the contributions of user-side and item-side contrastive learning is:
\begin{equation}
\mathcal{L}_{ssl} = \mathcal{L}_{ssl,user} + \lambda \cdot \mathcal{L}_{ssl,item}
\end{equation}
where $\lambda$ is a hyperparameter controlling the relative importance of the two loss terms.

\subsection{Playtime-guided Multimodal Random Walks}
To model users’ interest migration, we adopt a random walks strategy guided by multimodal similarity and playtime-informed patterns, ensuring alignment with core preferences. This approach enhances category coverage by leveraging multimodal embeddings that capture semantic associations beyond explicit category boundaries (see Section~\ref{sec:modal-cross-category}) and by introducing a category balance coefficient that encourages cross-category exploration.

\subsubsection{Playtime-based Random Selection (Initial node Selection)}
For each player $u$, we perform multiple rounds of random walks. At the beginning of each exploration round, we first select an initial exploration node from the user's strong interest game set $\mathcal{I}_u^{S}$. To reflect the user's interest intensity across different games, we use the normalized playtime $\hat{t}^u_i$ on game $i$ as the sampling weight and apply a multinomial sampling strategy. The probability of selecting game node $i$ is defined as follows:

\begin{equation}
P_u(i) = \frac{\hat{t}^u_i}{\sum_{j \in \mathcal{I}_u^{S}} \hat{t}^u_j}
\end{equation}

During multi-round exploration, we record the set of games $\mathcal{A}_u^i$ expanded from each initial node $i$ for user $u$. If node $i$ is selected again, a game from $\mathcal{A}_u^i$ is randomly chosen as a new starting point to maintain interest continuity and expand the exploration space, thereby improving category coverage and recommendation diversity. This progressive strategy avoids redundant paths and better utilizes recommendation opportunities. All added games form $\mathcal{A}_u = \bigcup_{i \in \mathcal{I}_u^{wi}} \mathcal{A}_u^i$, which is used to prevent repeated recommendations.

\subsubsection{Category-based Candidate Node Selection}
Starting from the initial game node $i$, we conduct hierarchical exploration. For the current node $i$, we first find the most modally similar game in each game category, and these most similar games from different categories form the candidate neighbor set $\Omega_u(i)$ of game $i$:
\begin{equation}
\Omega_u(i) = \bigcup_{c \in \mathcal{C}} \{j \mid j = \arg\max_{j \in \mathcal{I}_c, j \neq i} \phi(\mathbf{e}_i^{m}, \mathbf{e}_j^{m})\}
\end{equation}
where $\mathcal{C}$ is the set of all categories, $\mathcal{I}_c$ represents the set of games in category $c$, $\mathbf{e}_i^{m}$ represents the modal embedding vector of game $i$, which is extracted based on Section \ref{sec:modal-cross-category}. 

To prevent repeated recommendations, we perform preliminary filtering on the candidate set $\Omega_u(i)$ to obtain a new filtered candidate set $\Omega'_u(i)$:
\begin{equation}
\Omega'_u(i) = \{j \in \Omega_u(i) \mid j \notin \mathcal{I}_u \land j \notin \mathcal{A}_u \land |c^*(j)| < Q\}
\label{eq:filtered_candidates}
\end{equation}
where $|c^*(j)|$ denotes the number of times category $c^*(j)$ appears in the already recommended set $\mathcal{A}_u$, $c^*(j)$ represents the category that game $j$ stands for in the candidate generation process, and $Q$ represents the maximum number of game nodes that can be added per category for each user.

\subsubsection{Multi-metric Candidate Node Scoring}
For each game $j$ in the candidate set $\Omega'_u(i)$, we consider three key metrics to select game nodes to establish new connections with player $u$:

\paragraph{a) Multimodal Similarity}
Based on the cosine similarity of game modal content embedding vectors:
\begin{equation}
sim_{m}(i, j) = \phi(\mathbf{e}_i^{m}, \mathbf{e}_j^{m})
\end{equation}

Multimodal similarity captures content-level semantic connections between games by integrating textual descriptions and visual features. It enables the recommendation of content-related games that may differ in collaborative patterns, enhancing both relevance and the potential for novel discovery.

\paragraph{b) Interest Similarity}

Based on the similarity of players' interest intensity scores on two games, we define the interest similarity as:
\begin{equation}
sim_{t}(i, j) = 1 - \frac{1}{|\mathcal{U}_{i,j}|} \sum_{u \in \mathcal{U}_{i,j}} |\gamma_{u,i} - \gamma_{u,j}|
\end{equation}
where $\gamma_{u,i}$ denotes the interest intensity score of user $u$ for game $i$ (i.e., $\gamma_{u,i}^{(S)}$), and $\mathcal{U}_{i,j}$ denotes the set of users who have played both games $i$ and $j$. Interest similarity captures implicit associations between games by measuring consistency in user engagement patterns. Complementing content-based multimodal analysis, it reveals behaviorally similar games across categories, enhancing recommendation diversity while preserving alignment with user preferences.

\paragraph{c) Category Balance Coefficient}
It controls the balance of recommendation quantities across categories, enhancing diversity selection:
\begin{equation}
\omega_u(j) = \frac{Q - |c^*(j)|}{Q}
\label{eq:balance}
\end{equation}
The category balance coefficient adjusts selection probabilities during random walks to prevent over-concentration in a few categories, promoting broader category coverage. 

These three metrics are normalized in the candidate game set to obtain a comprehensive score:
\begin{equation}
score_u(i,j) = Agg(\hat{sim}_m(i, j)  ,  \hat{sim}_{t}(i, j) ,\hat{\omega}_u(c^*(j)))
\end{equation}
where $\hat{sim}_m(i,j)$, $\hat{sim}_{t}(i,j)$, and $\hat{\omega}_u(c^*(j))$ are the normalized multimodal similarity, interest similarity, and category balance coefficient, respectively. $Agg(\cdot)$ can be any differentiable function. We use equal-weight averaging to avoid tuning and ensure fairness. This fused scoring strategy integrates semantic relevance, behavioral patterns, and category diversity to balance recommendation accuracy and exploration. 

\subsubsection{Score-based Random Selection (Node Walking)}
Based on the comprehensive game scores, we apply multinomial sampling to the filtered candidate set $\Omega'_u(i)$ to select nodes to establish connections with player $u$, with the probability of selecting the next node being:
\begin{equation}
P^u_i(j) = \frac{score_u(i,j)}{\sum_{k \in \Omega'_u(i)} score_u(i,k)}
\end{equation}

This probabilistic strategy ensures relevance while promoting diversity among eligible recommendations.

\subsubsection{Multi-factor Path Weight Calculation}
To quantify the influence of explored nodes, we define a path-based weighting scheme. Given a path $i_1 \rightarrow i_2 \rightarrow \dots \rightarrow i_l$, the edge weight from user $u$ to node $i_l$ is computed as:

\begin{equation}
w^u_{i_l} =
\begin{cases}
\hat{t}^u_{i_1} & \text{if } l = 1 \\
\hat{t}^u_{i_1} \cdot sim_m(i_1, i_2) & \text{if } l = 2 \\
\hat{t}^u_{i_1} \cdot \prod_{k=1}^{l-1} sim_m(i_k, i_{k+1}) & \text{if } l \geq 3
\end{cases}
\end{equation}

This formulation combines the user's initial interest (playtime) with accumulated multimodal similarity along the path, balancing personalization and exploratory diversity during random walks. We adopt weighted LightGCN \cite{he2020lightgcn} to compute the final multimodal random walks embeddings. 
Incorporating our path weights, node embeddings are updated as:
\begin{equation}
\mathbf{e}_{u}^{(l)} = \sum_{i \in \mathcal{N}_u} \frac{w^u_i}{\sqrt{|\mathcal{N}_u||\mathcal{N}_i|}} \cdot \mathbf{e}_i^{(l-1)}, \mathbf{e}_{i}^{(l)} = \sum_{u \in \mathcal{N}_i} \frac{w^u_i}{\sqrt{|\mathcal{N}_i||\mathcal{N}_u|}} \cdot \mathbf{e}_u^{(l-1)}
\end{equation}
where $\mathcal{N}_u$ and $\mathcal{N}_i$ are the neighbor sets of user $u$ and game $i$ respectively, and $w^u_i$ is the path-based edge weight. This weighted aggregation emphasizes nodes with stronger semantic similarity and user interest. After $L$ propagation layers, the final user and item embeddings are defined as $\mathbf{e}_{u}^{MRW} = \mathbf{e}_{u}^{(L)}$, $\mathbf{e}_{i}^{MRW} = \mathbf{e}_{i}^{(L)}$. 
\subsubsection{Theoretical Analysis: Diversity Optimization via Controlled Martingale Process}

We model the MRW as a controlled stochastic process. Specifically, a user’s exploration path on the game graph is treated as a discrete-time random walk, where diversity at time step $t$ is defined by the number of distinct categories covered so far.

Let $P = \{i_1, i_2, \dots, i_t\}$ denote the set of game nodes visited by the user up to step $t$, and let each node $i$ belong to a set of categories $c(i) \subseteq \mathcal{C}$. We define the diversity metric at time $t$ as:
\begin{equation}
C_t(P) := \left| \bigcup_{j \in P} c(j) \right|
\label{eq:diversity}
\end{equation}

Let $\mathcal{F}_t$ represent the filtration containing all historical information up to time $t$. We model the sequence $\{C_t(P)\}_{t=0}^T$ as a controlled stochastic process. 

To encourage exploration into underrepresented categories, the transition probability $P^u_i(j)$ incorporates a category balance coefficient. The marginal contribution of a candidate node $j$ is quantified by its expected diversity gain, defined as:
\begin{equation}
\Delta C(j) := \mathbb{E}[C_t(P \cup \{j\}) - C_t(P) \mid \mathcal{F}_t]
\label{eq:delta}
\end{equation}

Here, $\mathbb{E}[\cdot \mid \mathcal{F}_t]$ denotes the conditional expectation given the current path history $\mathcal{F}_t$.

Given the definition of $C_t(P)$, it follows that:
\begin{equation}
\Delta C(j) = \bigl| c(j) \setminus c(P) \bigr|
\label{eq:delta_c_new}
\end{equation}
where $c(P) = \bigcup_{i \in P} c(i)$ denotes the set of categories already covered by the path $P$, and $\setminus$ denotes the set difference operator. $\Delta C(j)$ represents the number of new categories introduced by node $j$.

Substituting this into the expectation, we obtain:
\begin{equation}
\mathbb{E}[C_{t+1}(P) - C_t(P) \mid \mathcal{F}_t] = \sum_{j \in \Omega'_u(i)} P^u_i(j) \cdot \Delta C(j)
\label{eq:submartingale}
\end{equation}

This confirms that $\{C_t(P)\}$ forms a controlled submartingale \cite{williams1991probability}, indicating that the expected diversity is non-decreasing over time.
Now let $T$ denote the stopping time when the user's exploration terminates. By the optional stopping theorem, the expected diversity at time $T$ satisfies:
\begin{equation}
\mathbb{E}[C_T(P)] = \mathbb{E}[C_0(P)] + \mathbb{E}\left[\sum_{t=0}^{T-1} \sum_{j \in \Omega'_u(i)} P^u_i(j) \cdot \Delta C(j)\right]
\label{eq:stopping_exact}
\end{equation}
This provides an exact characterization of expected diversity gains, showing that MRW's category-aware mechanism systematically guides users to novel semantic regions. The node sequence $P$ forms a subgraph revealing behaviors like cross-category transitions, highlighting that diversity is structured, controllable graph traversal.

\subsection{Balance Module}

To integrate the above two modules, we introduce a balance module that dynamically controls their relative influence. This design enables flexible trade-offs between recommendation accuracy and diversity. We define the fused user and item embeddings as:
\begin{equation}
\mathbf{e}_{u} = \alpha \cdot \mathbf{e}_{u}^{IIE} + \mathbf{e}_{u}^{MRW}, \mathbf{e}_{i} = \alpha \cdot \mathbf{e}_{i}^{IIE} + \mathbf{e}_{i}^{MRW}
\end{equation}
where $\alpha > 0$ is a tunable parameter: a larger $\alpha$ emphasizes behavior-driven personalization (IIE), enhancing accuracy; a smaller $\alpha$ favors semantic-guided exploration (MRW), boosting diversity.

We introduce a balance-aware loss function inspired by CPGRec that adaptively reweights negative samples to balance accuracy and diversity:
\begin{equation}
\mathcal{L}_{\text{balance}} = - \sum_{(u, i, j)} \log \sigma \left( s_{u,i} - \tilde{s}_{u,j} \right)
\end{equation}
where $s_{u,i}=e_u\cdot e_i$ is the predicted score between user $u$ and positive item $i$, and $\tilde{s}_{u,j}$ is the reweighted score of the negative item $j$:
\begin{equation}
\tilde{s}_{u,j} = s_{u,j} \cdot \left( \frac{1}{1 + e^{-s_{u,j} \cdot \zeta}} \cdot K \right)
\end{equation}
where $\zeta$ controls the reweighting intensity, and $K$ is a scaling factor. This formulation suppresses easy negatives and highlights informative ones, enhancing the model's training efficiency.

\section{Experiments}
\subsection{Experimental Setup}
\subsubsection{Dataset}

To evaluate $DP^2Rec$, we use the Steam platform—the most common data source in game recommendation research due to its large user base, rich interaction data, and public APIs. We collected the Steam dataset following DRGame's method and crawled additional multimodal content (text and images). The filtered dataset contains 60,742 players, 7,726 games, 4,145,359 interactions, and 20 genres.

\subsubsection{Evaluation Metrics}

To evaluate recommendation quality, we measure accuracy using NDCG@K, Recall@K, Hit Ratio@K, and Precision@K. For diversity, we assess Coverage@K, which tracks the number of distinct game categories recommended. K is set to \{5, 10, 20\}. 
\subsubsection{Baselines}

For \textbf{accuracy}, we include non-modal SCGRec \cite{SCGRec} and RGCF \cite{tian2022learning}, and accuracy-oriented CPGRec\cite{li2024category}. We also consider multimodal models—BM3~\cite{zhou2023bootstrap}, SLMRec~\cite{tao2022self}, MGCN~\cite{yu2023multi}, and SMORE~\cite{ong2025spectrum}, which leverage visual and textual game content. For \textbf{diversity}, we compare $DP^2Rec$ with non-modal diversity-enhancing methods: EDUA\cite{liang2021enhancing}, DGCN\cite{zheng2021dgcn}, DGRec\cite{yang2023dgrec}, and diversity-focused CPGRec~\cite{li2024category}, as  existing multimodal models do not explicitly target diversity. To assess \textbf{accuracy-diversity trade-off}, we highlight comparisons with balance-driven CPGRec~\cite{li2024category}, which optimizes both objectives.

\subsection{Overall Performance}
\begin{table*}[ht]
\centering
\caption{Performance comparison of different recommendation methods.}
\label{table:performance}
\begin{adjustbox}{width=1.0\textwidth}
\begin{tabular}{ll|ccc|ccc|ccc|ccc|ccc}
\toprule
& \multirow{2}{*}{Method} & \multicolumn{3}{c}{NDCG} & \multicolumn{3}{c}{Recall} & \multicolumn{3}{c}{Hit Ratio} & \multicolumn{3}{c}{Precision} & \multicolumn{3}{c}{Coverage} \\
\cmidrule(lr){3-5} \cmidrule(lr){6-8} \cmidrule(lr){9-11} \cmidrule(lr){12-14} \cmidrule(lr){15-17}
& & @5 & @10 & @20 & @5 & @10 & @20 & @5 & @10 & @20 & @5 & @10 & @20 & @5 & @10 & @20 \\
\midrule
\multirow{8}{*}{\makecell[l]{Accuracy-based\\Methods}} 
& SCGRec & 0.2150 & 0.2283 & 0.2592 & 0.1341 & 0.2139 & 0.3208 & 0.6039 & 0.7663 & 0.8785 & 0.1867 & 0.1619 & 0.1331 & 3.2348 & 4.7899 & 6.5225 \\
& RGCF & 0.2628 & 0.2525 & 0.2607 & 0.1308 & 0.1901 & 0.2651 & 0.6423 & 0.7414 & 0.8179 & 0.2272 & 0.1822 & 0.1377 & 4.8120 & 6.7300 & \underline{8.5280} \\
& {BM3} & 0.2344 & 0.2291 & 0.2415 & 0.1296 & 0.1912 & 0.2662 & 0.6304 & 0.7561 & 0.8439 & 0.1933 & 0.1513 & 0.1144 & 5.1940 & \underline{6.8601} & \textbf{8.8693} \\
& {SLMRec} & 0.2621 & 0.2638 & 0.2841 & 0.1557 & \underline{0.2328} & \underline{0.3279} & 0.6780 & 0.8076 & \underline{0.8954} & 0.2205 & 0.1785 & 0.1359 & \textbf{5.4632} & \textbf{6.8632} & 8.3717 \\
& {MGCN} & 0.2615 & 0.2595 & 0.2777 & 0.1437 & 0.2163 & 0.3090 & 0.6739 & 0.7939 & 0.8720 & 0.2252 & 0.1822 & 0.1407 & 4.9905 & 6.6059 & 8.2074 \\
& SMORE & \underline{0.2741} & \underline{0.2720} & \underline{0.2903} & \underline{0.1558} & 0.2302 & 0.3242 & \underline{0.6970} & \underline{0.8112} & 0.8871 & \underline{0.2320} & \underline{0.1860} & \underline{0.1427} & 5.0582 & 6.5760 & 8.1871 \\
& Accuracy-focused CPGRec & 0.2594 & 0.2591 & 0.2791 & 0.1444 & 0.2213 & 0.3186 & 0.6675 & 0.7933 & 0.8783 & 0.2227 & 0.1812 & 0.1398 & 5.1467 & 6.6551 & 8.3357 \\
& Accuracy-focused $DP^2Rec$ & \textbf{0.3032} & \textbf{0.2978} & \textbf{0.3169} & \textbf{0.1704} & \textbf{0.2504} & \textbf{0.3503} & \textbf{0.7280} & \textbf{0.8342} & \textbf{0.9008} & \textbf{0.2570} & \textbf{0.2050} & \textbf{0.1555} & \underline{5.2076} & 6.6935 & 8.2847 \\
\midrule
\multirow{5}{*}{\makecell[l]{Diversity-based\\Methods}}
& EDUA & 0.1421 & 0.1455 & 0.1571 & 0.0908 & 0.1390 & 0.1897 & 0.4500 & 0.5700 & 0.6600 & 0.1200 & 0.0935 & 0.0685 & 5.6850 & 7.4700 & 9.1200 \\
& DGCN & 0.0963 & 0.0900 & 0.0799 & 0.0601 & 0.1031 & 0.1689 & 0.3668 & 0.5327 & 0.6925 & 0.0923 & 0.0850 & 0.0732 & 5.3660 & 7.4756 & 9.2498 \\
& DGRec & 0.1134 & 0.1075 & 0.0931 & 0.0959 & 0.1529 & 0.2165 & 0.4988 & 0.6660 & 0.7860 & 0.1298 & 0.1128 & 0.0896 & 5.8438 & \underline{7.9892} & \underline{9.9746} \\
& Diversity-focused CPGRec & \underline{0.2257} & \underline{0.2203} & \underline{0.2343} & \underline{0.1208} & \underline{0.1851} & \underline{0.2671} & \underline{0.6076} & \underline{0.7376} & \underline{0.8341} & \underline{0.1941} & \underline{0.1544} & \underline{0.1163} & \underline{6.0860} & 7.6519 & 9.3547 \\
& Diversity-focused $DP^2Rec$ & \textbf{0.2346} & \textbf{0.2322} & \textbf{0.2518} & \textbf{0.1273} & \textbf{0.1942} & \textbf{0.2862} & \textbf{0.6298} & \textbf{0.7626} & \textbf{0.8586} & \textbf{0.1953} & \textbf{0.1577} & \textbf{0.1241} & \textbf{7.9133} & \textbf{11.2227} & \textbf{14.7972} \\
\midrule
\multirow{2}{*}{\makecell[l]{Balance-based\\Methods}}
& CPGRec (trade-off model) & \underline{0.2347} & \underline{0.2333} & \underline{0.2532} & \underline{0.1272} & \underline{0.1952} & \underline{0.2893} & \underline{0.6326} & \underline{0.7557} & \underline{0.8512} & \underline{0.2027} & \underline{0.1666} & \underline{0.1308} & \underline{5.6811} & \underline{7.0266} & \underline{8.4586} \\
& $DP^2Rec$ (trade-off model)& \textbf{0.2518} & \textbf{0.2497} & \textbf{0.2688} & \textbf{0.1403} & \textbf{0.2100} & \textbf{0.3038} & \textbf{0.6587} & \textbf{0.7846} & \textbf{0.8722} & \textbf{0.2121} & \textbf{0.1712} & \textbf{0.1330} & \textbf{6.6499} & \textbf{9.3627} & \textbf{12.0228} \\
\bottomrule
\end{tabular}
\end{adjustbox}
\label{tab:comparison} 
\end{table*}

\begin{table*}[t]
\centering
\caption{Ablation study. We show $DP^2Rec$’s performance when removing each of the submodules.}
\resizebox{\textwidth}{!}{
\begin{tabular}{lccccccccccccccc}
\toprule
\multirow{2}{*}{Method} & \multicolumn{3}{c}{NDCG} & \multicolumn{3}{c}{Recall} & \multicolumn{3}{c}{Hit Ratio} & \multicolumn{3}{c}{Precision} & \multicolumn{3}{c}{Coverage} \\
\cmidrule(lr){2-4} \cmidrule(lr){5-7} \cmidrule(lr){8-10} \cmidrule(lr){11-13} \cmidrule(lr){14-16}
 & @5 & @10 & @20 & @5 & @10 & @20 & @5 & @10 & @20 & @5 & @10 & @20 & @5 & @10 & @20 \\
\midrule
\textit{Accuracy-focused $DP^2Rec$} & \textbf{0.3032} & \textbf{0.2978} & \textbf{0.3169} & \textbf{0.1704} & \textbf{0.2504} & \textbf{0.3503} & \textbf{0.7280} & \textbf{0.8342} & \textbf{0.9008} & \textbf{0.2570} & \textbf{0.2050} & \textbf{0.1555} & \textbf{5.2076} & \textbf{6.6935} & \textbf{8.2847} \\
\quad w/o Dual-Interest Recognition via EM & 0.2877 & 0.2844 & 0.3030 & 0.1609 & 0.2395 & 0.3381 & 0.7069 & 0.8190 & 0.8918 & 0.2446 & 0.1962 & 0.1496 & 5.0780 & 6.5365 & 8.1875 \\
\quad w/o Dual-Interest Graph Learning & 0.2743 & 0.2694 & 0.2848 & 0.1454 & 0.2180 & 0.3083 & 0.6785 & 0.7841 & 0.8561 & 0.2370 & 0.1914 & 0.1456 & 4.8388 & 6.2882 & 7.9043 \\
\midrule
\textit{Diversity-focused $DP^2Rec$} & 0.2346 & 0.2322 & 0.2518 & 0.1273 & 0.1942 & 0.2862 & 0.6298 & 0.7626 & 0.8586 & 0.1953 & 0.1577 & 0.1241 & \textbf{7.9133} & \textbf{11.2227} & \textbf{14.7972} \\
\quad w/o Category-based Candidate Node Selection& \textbf{0.2528} & \textbf{0.2506} & \textbf{0.2678} & \textbf{0.1403} & \textbf{0.2120} & \textbf{0.3011} & \textbf{0.6606} & \textbf{0.7848} & \textbf{0.8705} & \textbf{0.2122} & \textbf{0.1703} & \textbf{0.1306} & 5.6348 & 6.9416 & 8.5220 \\
\quad w/o Category Balance Coefficient & 0.2493 & 0.2472 & 0.2629 & 0.1388 & 0.2095 & 0.2952 & 0.6572 & 0.7817 & 0.8653 & 0.2101 & 0.1686 & 0.1282 & 5.9722 & 7.4482 & 11.2468 \\
\bottomrule
\end{tabular}
}
\label{table:ablation} 
\end{table*}
\subsubsection{Performance Evaluation}

We evaluated our proposed model, $DP^2Rec$, against several mainstream baseline models. The evaluation results are presented in Table \ref{table:performance}, where the best and second-best results for each metric are highlighted in \textbf{bold} and \underline{underlined}, respectively. Each result represents the average over five runs. Based on the results, we summarize the following key conclusions: 

\begin{itemize}

    \item \textbf{Advantages of $DP^2Rec$.}  
    $DP^2Rec$ outperforms diversity-based baselines, surpassing DGRec in \textbf{Coverage} by 13–21\%, while maintaining competitive \textbf{NDCG} and \textbf{Hit Ratio} with the SOTA accuracy-focused method in game recommendation (RGCF). This shows it balances diversity and accuracy effectively.

    \item \textbf{Superiority on Dual Objectives.}  
    The accuracy-focused $DP^2Rec$ surpasses the best non-modal (CPGRec) and multimodal (SMORE) baselines by ~17\% and ~11\% on NDCG@5, respectively. Meanwhile, its diversity-focused variant exceeds the leading diversity model (DGRec) in Coverage by 30–50\%, showing strong performance across both objectives.

    \item \textbf{Inherent Trade-off Between Accuracy and Diversity.}  
    While SMORE leads among current accuracy-oriented models (excluding the accuracy-optimized $DP^2Rec$), it falls short on Coverage compared to diversity-focused models. This highlights the inherent trade-off—pursuing accuracy often comes at the cost of diversity, which can reduce recommendation variety.

\end{itemize}

\subsubsection{Ablation Study}

We perform ablation experiments to evaluate the contribution of four core submodules: (1) Dual-Interest Recognition via EM, (2) Dual-Interest Graph Learning, (3) Category-based Candidate Node Selection, and (4) Category Balance Coefficient. The corresponding variants are denoted as \textit{w/o DIR}, \textit{w/o DIGL}, \textit{w/o CCNS}, and \textit{w/o CBC}. Results in Table~\ref{table:ablation} offer the following insights:

\textbf{1) \textit{w/o DIR.}}
Without the EM-based dual-beta modeling, strong and weak interests remain entangled, limiting the model's capacity to capture true two-level user preferences and reducing accuracy. Additionally, the inability to leverage weak interests for cross-category exploration results in diminished diversity.

\textbf{2) \textit{w/o DIGL.}}
Removing contrastive learning causes the model to focus solely on strong interests, reducing both accuracy and diversity. This over-reliance overlooks weak interests, which not only contribute to finer-grained preference modeling but also promote exploration across categories.

\textbf{3) \textit{w/o CCNS.}}
This submodule enables cross-category exploration via semantic similarity. Its removal increases accuracy slightly, by focusing on familiar regions, but significantly reduces diversity, confirming its importance in discovering semantically relevant games outside core categories.

\textbf{4) \textit{w/o CBC.}} 
Dropping this submodule leads to more recommendations from dominant categories, improving accuracy but sharply reducing diversity. This confirms its role in ensuring balanced exposure across categories and preventing over-concentration.

\subsubsection{Parameter Sensitivity}

\paragraph{Sensitivity Analysis on $\alpha$}

\begin{figure}[ht]
    \begin{subfigure}{0.48\linewidth}
        \centering
        \includegraphics[width=\linewidth]{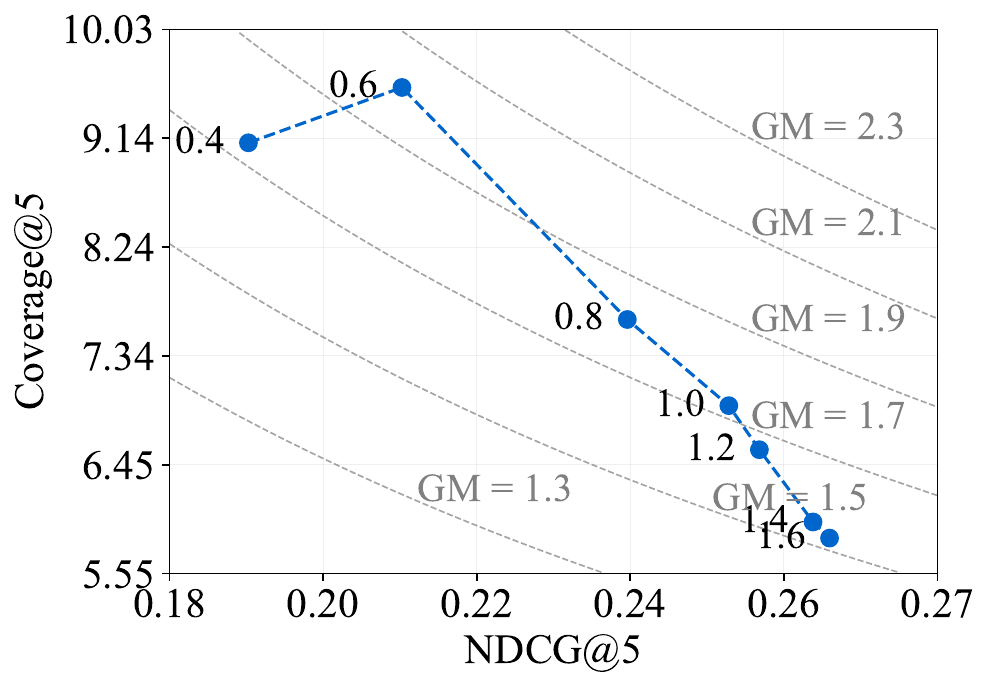}
        \caption{Analysis of $\alpha$}
        \label{fig:alpha_tradeoff}
    \end{subfigure}
    \begin{subfigure}{0.48\linewidth}
        \centering
        \includegraphics[width=\linewidth]{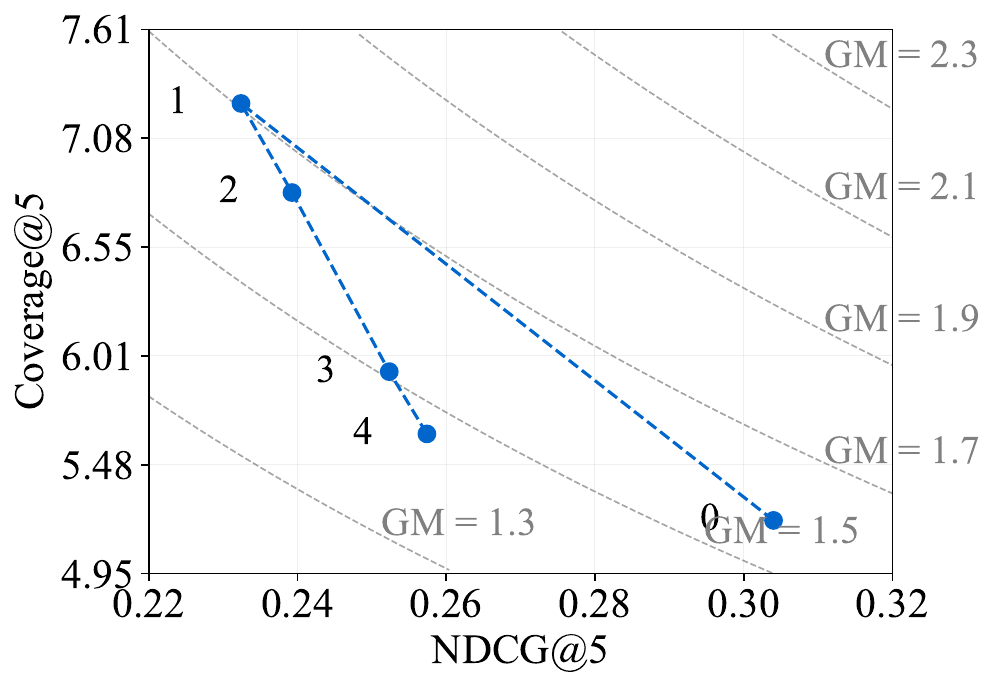}
        \caption{Analysis of  $Q$}
        \label{fig:q_tradeoff}
    \end{subfigure}
    \caption{Analysis of key hyperparameters $\alpha$ and $Q$.}
    \label{fig:param_analysis}
\end{figure}

As shown in Figure~\ref{fig:param_analysis}(a), increasing $\alpha$ from 0.4 to 1.6 steadily boosts NDCG@5 (from 0.19 to 0.26), reflecting improved capture of users’ core preferences. However, this gain in accuracy comes with a drop in diversity, as Coverage@5 declines from 9.10 to 5.84—indicating a narrowing of recommendation scope. To quantify this trade-off, we compute the \textbf{geometric mean (GM)} of NDCG@5 and Coverage@5 ($GM = \sqrt{\text{NDCG@5} \times \text{Coverage@5}}$). The contour map reveals that the optimal balance occurs at $\alpha = 0.6$, where GM peaks at \textbf{2.00}.

\paragraph{Sensitivity Analysis on $Q$}

As shown in Figure~\ref{fig:param_analysis}(b), increasing $Q$ from 0 to 4 causes NDCG@5 to initially drop from 0.30 to 0.24 before gradually recovering to 0.26. Meanwhile, Coverage@5 peaks at $Q = 1$ (7.25) and then slightly declines. This suggests that moderate category control enhances structural diversity, while excessive flexibility narrows recommendations.
To evaluate the overall performance, we compute the GM of NDCG@5 and Coverage@5. The contour plot shows the highest GM at $Q = 1$, reaching \textbf{1.71}, indicating it offers the best balance between accuracy and diversity.

\section{Conclusion}
In this work, we address two core challenges in game recommendation: the underutilization of playtime as a behavioral signal and the limited capacity to promote cross-category diversity based on multimodal similarity. To enhance accuracy, we introduce a \textit{IIE} module that distinguishes strong and weak interests based on playtime distributions. To promote diversity, we design a \textit{MRW} that incorporates both behavioral and semantic signals to guide user exploration across game categories.

\begin{acks}
This work was partially supported by
the \grantsponsor{NSFC}{National Natural Science Foundation of China}{https://www.nsfc.gov.cn/}
under Project No.~\grantnum{NSFC}{62202122} and No.~\grantnum{NSFC}{62073272};
by the \grantsponsor{GDBARF}{Guangdong Basic and Applied Basic Research Foundation}{http://gdstc.gd.gov.cn/}
under Grant No.~\grantnum{GDBARF}{2024A1515011949};
by the \grantsponsor{SZSTP}{Shenzhen Science and Technology Program}{http://stic.sz.gov.cn/}
under Grant No.~\grantnum{SZSTP}{GXWD20231130110308001}, \grantnum{SZSTP}{JCYJ20240813104843058}, and \grantnum{SZSTP}{JCYJ20240813104837050};
by the \grantsponsor{SZES}{Shenzhen Education Science ``14th Five-Year Plan'' 2023 Annual Project on Artificial Intelligence Special Project}{http://szeb.sz.gov.cn/}
under Grant No.~\grantnum{SZES}{rgzn23001};
by the \grantsponsor{GPEHR}{Guangdong Province Higher Education Research and Reform Project}{http://edu.gd.gov.cn/}
under Grant No.~\grantnum{GPEHR}{YueJiaoGaoHan(2024) No. 9};
and by the \grantsponsor{SZKLIIC}{Shenzhen Key Laboratory of Internet Information Collaboration}{http://www.sz.gov.cn/cn/xxgk/zfxxgj/zdsys/}
under Grant No.~\grantnum{SZKLIIC}{ZDSY20120613125016389}.
\end{acks}
\bibliographystyle{ACM-Reference-Format}
\bibliography{reference}


\begin{thebibliography}{55}


\ifx \showCODEN    \undefined \def \showCODEN     #1{\unskip}     \fi
\ifx \showISBNx    \undefined \def \showISBNx     #1{\unskip}     \fi
\ifx \showISBNxiii \undefined \def \showISBNxiii  #1{\unskip}     \fi
\ifx \showISSN     \undefined \def \showISSN      #1{\unskip}     \fi
\ifx \showLCCN     \undefined \def \showLCCN      #1{\unskip}     \fi
\ifx \shownote     \undefined \def \shownote      #1{#1}          \fi
\ifx \showarticletitle \undefined \def \showarticletitle #1{#1}   \fi
\ifx \showURL      \undefined \def \showURL       {\relax}        \fi
\providecommand\bibfield[2]{#2}
\providecommand\bibinfo[2]{#2}
\providecommand\natexlab[1]{#1}
\providecommand\showeprint[2][]{arXiv:#2}

\bibitem[Abdool et~al\mbox{.}(2020)]%
        {abdool2020managing}
\bibfield{author}{\bibinfo{person}{Mustafa Abdool}, \bibinfo{person}{Malay Haldar}, \bibinfo{person}{Prashant Ramanathan}, \bibinfo{person}{Tyler Sax}, \bibinfo{person}{Lanbo Zhang}, \bibinfo{person}{Aamir Manaswala}, \bibinfo{person}{Lynn Yang}, \bibinfo{person}{Bradley Turnbull}, \bibinfo{person}{Qing Zhang}, {and} \bibinfo{person}{Thomas Legrand}.} \bibinfo{year}{2020}\natexlab{}.
\newblock \showarticletitle{Managing diversity in airbnb search}. In \bibinfo{booktitle}{\emph{Proceedings of the 26th ACM SIGKDD International Conference on Knowledge Discovery \& Data Mining}}. \bibinfo{pages}{2952--2960}.
\newblock


\bibitem[Anwar et~al\mbox{.}(2017)]%
        {anwar2017game}
\bibfield{author}{\bibinfo{person}{Syed~Muhammad Anwar}, \bibinfo{person}{Talha Shahzad}, \bibinfo{person}{Zunaira Sattar}, \bibinfo{person}{Rahma Khan}, {and} \bibinfo{person}{Muhammad Majid}.} \bibinfo{year}{2017}\natexlab{}.
\newblock \showarticletitle{A game recommender system using collaborative filtering (GAMBIT)}. In \bibinfo{booktitle}{\emph{2017 14th International Bhurban Conference on Applied Sciences and Technology (IBCAST)}}. IEEE, \bibinfo{pages}{328--332}.
\newblock


\bibitem[Bai et~al\mbox{.}(2025a)]%
        {bai2025qwen2}
\bibfield{author}{\bibinfo{person}{Shuai Bai}, \bibinfo{person}{Keqin Chen}, \bibinfo{person}{Xuejing Liu}, \bibinfo{person}{Jialin Wang}, \bibinfo{person}{Wenbin Ge}, \bibinfo{person}{Sibo Song}, \bibinfo{person}{Kai Dang}, \bibinfo{person}{Peng Wang}, \bibinfo{person}{Shijie Wang}, \bibinfo{person}{Jun Tang}, {et~al\mbox{.}}} \bibinfo{year}{2025}\natexlab{a}.
\newblock \showarticletitle{Qwen2. 5-vl technical report}.
\newblock \bibinfo{journal}{\emph{arXiv preprint arXiv:2502.13923}} (\bibinfo{year}{2025}).
\newblock


\bibitem[Bai et~al\mbox{.}(2025b)]%
        {bai2025chime}
\bibfield{author}{\bibinfo{person}{Yong Bai}, \bibinfo{person}{Rui Xiang}, \bibinfo{person}{Kaiyuan Li}, \bibinfo{person}{Yongxiang Tang}, \bibinfo{person}{Yanhua Cheng}, \bibinfo{person}{Xialong Liu}, \bibinfo{person}{Peng Jiang}, {and} \bibinfo{person}{Kun Gai}.} \bibinfo{year}{2025}\natexlab{b}.
\newblock \showarticletitle{CHIME: A Compressive Framework for Holistic Interest Modeling}.
\newblock \bibinfo{journal}{\emph{arXiv preprint arXiv:2504.06780}} (\bibinfo{year}{2025}).
\newblock


\bibitem[BharathiPriya et~al\mbox{.}(2021)]%
        {bharathipriya2021online}
\bibfield{author}{\bibinfo{person}{C BharathiPriya}, \bibinfo{person}{Akash Sreenivasu}, {and} \bibinfo{person}{Sampath Kumar}.} \bibinfo{year}{2021}\natexlab{}.
\newblock \showarticletitle{Online Video Game Recommendation System Using Content And Collaborative Filtering Techniques}. In \bibinfo{booktitle}{\emph{2021 International Conference on Advancements in Electrical, Electronics, Communication, Computing and Automation (ICAECA)}}. IEEE, \bibinfo{pages}{1--7}.
\newblock


\bibitem[Carbonell and Goldstein(1998)]%
        {carbonell1998use}
\bibfield{author}{\bibinfo{person}{Jaime Carbonell} {and} \bibinfo{person}{Jade Goldstein}.} \bibinfo{year}{1998}\natexlab{}.
\newblock \showarticletitle{The use of MMR, diversity-based reranking for reordering documents and producing summaries}. In \bibinfo{booktitle}{\emph{Proceedings of the 21st annual international ACM SIGIR conference on Research and development in information retrieval}}. \bibinfo{pages}{335--336}.
\newblock


\bibitem[Caroux(2023)]%
        {caroux2023presence}
\bibfield{author}{\bibinfo{person}{Lo{\"\i}c Caroux}.} \bibinfo{year}{2023}\natexlab{}.
\newblock \showarticletitle{Presence in video games: A systematic review and meta-analysis of the effects of game design choices}.
\newblock \bibinfo{journal}{\emph{Applied Ergonomics}}  \bibinfo{volume}{107} (\bibinfo{year}{2023}), \bibinfo{pages}{103936}.
\newblock


\bibitem[Chen et~al\mbox{.}(2024b)]%
        {chen2024post}
\bibfield{author}{\bibinfo{person}{Chaochao Chen}, \bibinfo{person}{Yizhao Zhang}, \bibinfo{person}{Yuyuan Li}, \bibinfo{person}{Jun Wang}, \bibinfo{person}{Lianyong Qi}, \bibinfo{person}{Xiaolong Xu}, \bibinfo{person}{Xiaolin Zheng}, {and} \bibinfo{person}{Jianwei Yin}.} \bibinfo{year}{2024}\natexlab{b}.
\newblock \showarticletitle{Post-training attribute unlearning in recommender systems}.
\newblock \bibinfo{journal}{\emph{ACM Transactions on Information Systems}} \bibinfo{volume}{43}, \bibinfo{number}{1} (\bibinfo{year}{2024}), \bibinfo{pages}{1--28}.
\newblock


\bibitem[Chen et~al\mbox{.}(2024a)]%
        {chen2024dlcrec}
\bibfield{author}{\bibinfo{person}{Jiaju Chen}, \bibinfo{person}{Chongming Gao}, \bibinfo{person}{Shuai Yuan}, \bibinfo{person}{Shuchang Liu}, \bibinfo{person}{Qingpeng Cai}, {and} \bibinfo{person}{Peng Jiang}.} \bibinfo{year}{2024}\natexlab{a}.
\newblock \showarticletitle{DLCRec: A Novel Approach for Managing Diversity in LLM-Based Recommender Systems}.
\newblock \bibinfo{journal}{\emph{arXiv preprint arXiv:2408.12470}} (\bibinfo{year}{2024}).
\newblock


\bibitem[Chen et~al\mbox{.}(2025)]%
        {chen2025dlcrec}
\bibfield{author}{\bibinfo{person}{Jiaju Chen}, \bibinfo{person}{Chongming Gao}, \bibinfo{person}{Shuai Yuan}, \bibinfo{person}{Shuchang Liu}, \bibinfo{person}{Qingpeng Cai}, {and} \bibinfo{person}{Peng Jiang}.} \bibinfo{year}{2025}\natexlab{}.
\newblock \showarticletitle{DLCRec: A Novel Approach for Managing Diversity in LLM-Based Recommender Systems}. In \bibinfo{booktitle}{\emph{Proceedings of the Eighteenth ACM International Conference on Web Search and Data Mining}}. \bibinfo{pages}{857--865}.
\newblock


\bibitem[Chen et~al\mbox{.}(2018)]%
        {chen2018fast}
\bibfield{author}{\bibinfo{person}{Laming Chen}, \bibinfo{person}{Guoxin Zhang}, {and} \bibinfo{person}{Eric Zhou}.} \bibinfo{year}{2018}\natexlab{}.
\newblock \showarticletitle{Fast greedy map inference for determinantal point process to improve recommendation diversity}.
\newblock \bibinfo{journal}{\emph{Advances in Neural Information Processing Systems}}  \bibinfo{volume}{31} (\bibinfo{year}{2018}).
\newblock


\bibitem[Cheuque et~al\mbox{.}(2019)]%
        {cheuque2019recommender}
\bibfield{author}{\bibinfo{person}{Germ{\'a}n Cheuque}, \bibinfo{person}{Jos{\'e} Guzm{\'a}n}, {and} \bibinfo{person}{Denis Parra}.} \bibinfo{year}{2019}\natexlab{}.
\newblock \showarticletitle{Recommender systems for online video game platforms: The case of steam}. In \bibinfo{booktitle}{\emph{Companion Proceedings of The 2019 World Wide Web Conference}}. \bibinfo{pages}{763--771}.
\newblock


\bibitem[Cho et~al\mbox{.}(2023)]%
        {cho2023dynamic}
\bibfield{author}{\bibinfo{person}{Junsu Cho}, \bibinfo{person}{Dongmin Hyun}, \bibinfo{person}{Dong won Lim}, \bibinfo{person}{Hyeon jae Cheon}, \bibinfo{person}{Hyoung-iel Park}, {and} \bibinfo{person}{Hwanjo Yu}.} \bibinfo{year}{2023}\natexlab{}.
\newblock \showarticletitle{Dynamic multi-behavior sequence modeling for next item recommendation}. In \bibinfo{booktitle}{\emph{Proceedings of the AAAI conference on artificial intelligence}}, Vol.~\bibinfo{volume}{37}. \bibinfo{pages}{4199--4207}.
\newblock


\bibitem[Coppolillo et~al\mbox{.}(2024)]%
        {coppolillo2024relevance}
\bibfield{author}{\bibinfo{person}{Erica Coppolillo}, \bibinfo{person}{Giuseppe Manco}, {and} \bibinfo{person}{Aristides Gionis}.} \bibinfo{year}{2024}\natexlab{}.
\newblock \showarticletitle{Relevance meets diversity: A user-centric framework for knowledge exploration through recommendations}. In \bibinfo{booktitle}{\emph{Proceedings of the 30th ACM SIGKDD Conference on Knowledge Discovery and Data Mining}}. \bibinfo{pages}{490--501}.
\newblock


\bibitem[Dempster et~al\mbox{.}(1977)]%
        {dempster1977maximum}
\bibfield{author}{\bibinfo{person}{Arthur~P Dempster}, \bibinfo{person}{Nan~M Laird}, {and} \bibinfo{person}{Donald~B Rubin}.} \bibinfo{year}{1977}\natexlab{}.
\newblock \showarticletitle{Maximum likelihood from incomplete data via the EM algorithm}.
\newblock \bibinfo{journal}{\emph{Journal of the royal statistical society: series B (methodological)}} \bibinfo{volume}{39}, \bibinfo{number}{1} (\bibinfo{year}{1977}), \bibinfo{pages}{1--22}.
\newblock


\bibitem[Devlin et~al\mbox{.}(2019)]%
        {devlin2019bert}
\bibfield{author}{\bibinfo{person}{Jacob Devlin}, \bibinfo{person}{Ming-Wei Chang}, \bibinfo{person}{Kenton Lee}, {and} \bibinfo{person}{Kristina Toutanova}.} \bibinfo{year}{2019}\natexlab{}.
\newblock \showarticletitle{Bert: Pre-training of deep bidirectional transformers for language understanding}. In \bibinfo{booktitle}{\emph{Proceedings of the 2019 conference of the North American chapter of the association for computational linguistics: human language technologies, volume 1 (long and short papers)}}. \bibinfo{pages}{4171--4186}.
\newblock


\bibitem[Duricic et~al\mbox{.}(2023)]%
        {duricic2023beyond}
\bibfield{author}{\bibinfo{person}{Tomislav Duricic}, \bibinfo{person}{Dominik Kowald}, \bibinfo{person}{Emanuel Lacic}, {and} \bibinfo{person}{Elisabeth Lex}.} \bibinfo{year}{2023}\natexlab{}.
\newblock \showarticletitle{Beyond-accuracy: a review on diversity, serendipity, and fairness in recommender systems based on graph neural networks}.
\newblock \bibinfo{journal}{\emph{Frontiers in big data}}  \bibinfo{volume}{6} (\bibinfo{year}{2023}), \bibinfo{pages}{1251072}.
\newblock


\bibitem[Eskandanian and Mobasher(2020)]%
        {eskandanian2020using}
\bibfield{author}{\bibinfo{person}{Farzad Eskandanian} {and} \bibinfo{person}{Bamshad Mobasher}.} \bibinfo{year}{2020}\natexlab{}.
\newblock \showarticletitle{Using stable matching to optimize the balance between accuracy and diversity in recommendation}. In \bibinfo{booktitle}{\emph{Proceedings of the 28th ACM Conference on User Modeling, Adaptation and Personalization}}. \bibinfo{pages}{71--79}.
\newblock


\bibitem[Gan et~al\mbox{.}(2020)]%
        {gan2020enhancing}
\bibfield{author}{\bibinfo{person}{Lu Gan}, \bibinfo{person}{Diana Nurbakova}, \bibinfo{person}{L{\'e}a Laporte}, {and} \bibinfo{person}{Sylvie Calabretto}.} \bibinfo{year}{2020}\natexlab{}.
\newblock \showarticletitle{Enhancing recommendation diversity using determinantal point processes on knowledge graphs}. In \bibinfo{booktitle}{\emph{Proceedings of the 43rd International ACM SIGIR Conference on Research and Development in Information Retrieval}}. \bibinfo{pages}{2001--2004}.
\newblock


\bibitem[He et~al\mbox{.}(2020)]%
        {he2020lightgcn}
\bibfield{author}{\bibinfo{person}{Xiangnan He}, \bibinfo{person}{Kuan Deng}, \bibinfo{person}{Xiang Wang}, \bibinfo{person}{Yan Li}, \bibinfo{person}{Yongdong Zhang}, {and} \bibinfo{person}{Meng Wang}.} \bibinfo{year}{2020}\natexlab{}.
\newblock \showarticletitle{Lightgcn: Simplifying and powering graph convolution network for recommendation}. In \bibinfo{booktitle}{\emph{Proceedings of the 43rd International ACM SIGIR conference on research and development in Information Retrieval}}. \bibinfo{pages}{639--648}.
\newblock


\bibitem[Huang et~al\mbox{.}(2021)]%
        {huang2021sliding}
\bibfield{author}{\bibinfo{person}{Yanhua Huang}, \bibinfo{person}{Weikun Wang}, \bibinfo{person}{Lei Zhang}, {and} \bibinfo{person}{Ruiwen Xu}.} \bibinfo{year}{2021}\natexlab{}.
\newblock \showarticletitle{Sliding spectrum decomposition for diversified recommendation}. In \bibinfo{booktitle}{\emph{Proceedings of the 27th ACM SIGKDD conference on knowledge discovery \& data mining}}. \bibinfo{pages}{3041--3049}.
\newblock


\bibitem[Ikram and Farooq(2022)]%
        {ikram2022multimedia}
\bibfield{author}{\bibinfo{person}{Fasiha Ikram} {and} \bibinfo{person}{Humera Farooq}.} \bibinfo{year}{2022}\natexlab{}.
\newblock \showarticletitle{Multimedia Recommendation System for Video Game Based on High-Level Visual Semantic Features}.
\newblock \bibinfo{journal}{\emph{Scientific Programming}} \bibinfo{volume}{2022}, \bibinfo{number}{1} (\bibinfo{year}{2022}), \bibinfo{pages}{6084363}.
\newblock


\bibitem[Li et~al\mbox{.}(2024)]%
        {li2024category}
\bibfield{author}{\bibinfo{person}{Xiping Li}, \bibinfo{person}{Jianghong Ma}, \bibinfo{person}{Kangzhe Liu}, \bibinfo{person}{Shanshan Feng}, \bibinfo{person}{Haijun Zhang}, {and} \bibinfo{person}{Yutong Wang}.} \bibinfo{year}{2024}\natexlab{}.
\newblock \showarticletitle{Category-based and Popularity-guided Video Game Recommendation: A Balance-oriented Framework}. In \bibinfo{booktitle}{\emph{Proceedings of the ACM on Web Conference 2024}}. \bibinfo{pages}{3734--3744}.
\newblock


\bibitem[Liang et~al\mbox{.}(2021)]%
        {liang2021enhancing}
\bibfield{author}{\bibinfo{person}{Yile Liang}, \bibinfo{person}{Tieyun Qian}, \bibinfo{person}{Qing Li}, {and} \bibinfo{person}{Hongzhi Yin}.} \bibinfo{year}{2021}\natexlab{}.
\newblock \showarticletitle{Enhancing domain-level and user-level adaptivity in diversified recommendation}. In \bibinfo{booktitle}{\emph{Proceedings of the 44th International ACM SIGIR Conference on Research and Development in Information Retrieval}}. \bibinfo{pages}{747--756}.
\newblock


\bibitem[Lin et~al\mbox{.}(2022)]%
        {lin2022feature}
\bibfield{author}{\bibinfo{person}{Zihan Lin}, \bibinfo{person}{Hui Wang}, \bibinfo{person}{Jingshu Mao}, \bibinfo{person}{Wayne~Xin Zhao}, \bibinfo{person}{Cheng Wang}, \bibinfo{person}{Peng Jiang}, {and} \bibinfo{person}{Ji-Rong Wen}.} \bibinfo{year}{2022}\natexlab{}.
\newblock \showarticletitle{Feature-aware diversified re-ranking with disentangled representations for relevant recommendation}. In \bibinfo{booktitle}{\emph{Proceedings of the 28th ACM SIGKDD Conference on Knowledge Discovery and Data Mining}}. \bibinfo{pages}{3327--3335}.
\newblock


\bibitem[Liu et~al\mbox{.}(2023b)]%
        {liu2023information}
\bibfield{author}{\bibinfo{person}{Chengliang Liu}, \bibinfo{person}{Jie Wen}, \bibinfo{person}{Zhihao Wu}, \bibinfo{person}{Xiaoling Luo}, \bibinfo{person}{Chao Huang}, {and} \bibinfo{person}{Yong Xu}.} \bibinfo{year}{2023}\natexlab{b}.
\newblock \showarticletitle{Information recovery-driven deep incomplete multiview clustering network}.
\newblock \bibinfo{journal}{\emph{IEEE Transactions on Neural Networks and Learning Systems}} \bibinfo{volume}{35}, \bibinfo{number}{11} (\bibinfo{year}{2023}), \bibinfo{pages}{15442--15452}.
\newblock


\bibitem[Liu et~al\mbox{.}(2023a)]%
        {liu2023modeling}
\bibfield{author}{\bibinfo{person}{Haobing Liu}, \bibinfo{person}{Jianyu Ding}, \bibinfo{person}{Yanmin Zhu}, \bibinfo{person}{Feilong Tang}, \bibinfo{person}{Jiadi Yu}, \bibinfo{person}{Ruobing Jiang}, {and} \bibinfo{person}{Zhongwen Guo}.} \bibinfo{year}{2023}\natexlab{a}.
\newblock \showarticletitle{Modeling multi-aspect preferences and intents for multi-behavioral sequential recommendation}.
\newblock \bibinfo{journal}{\emph{Knowledge-Based Systems}}  \bibinfo{volume}{280} (\bibinfo{year}{2023}), \bibinfo{pages}{111013}.
\newblock


\bibitem[Liu et~al\mbox{.}(2024)]%
        {liu2024drgame}
\bibfield{author}{\bibinfo{person}{Kangzhe Liu}, \bibinfo{person}{Jianghong Ma}, \bibinfo{person}{Shanshan Feng}, \bibinfo{person}{Haijun Zhang}, {and} \bibinfo{person}{Zhao Zhang}.} \bibinfo{year}{2024}\natexlab{}.
\newblock \showarticletitle{DRGame: Diversified Recommendation for Multi-category Video Games with Balanced Implicit Preferences}. In \bibinfo{booktitle}{\emph{International Conference on Database Systems for Advanced Applications}}. Springer, \bibinfo{pages}{254--263}.
\newblock


\bibitem[Lv et~al\mbox{.}(2025)]%
        {lv2025dynamic}
\bibfield{author}{\bibinfo{person}{Mingyang Lv}, \bibinfo{person}{Xiangfeng Liu}, {and} \bibinfo{person}{Yuanbo Xu}.} \bibinfo{year}{2025}\natexlab{}.
\newblock \showarticletitle{Dynamic Multi-Interest Graph Neural Network for Session-Based Recommendation}.
\newblock  (\bibinfo{year}{2025}).
\newblock


\bibitem[Massey~Jr(1951)]%
        {massey1951kolmogorov}
\bibfield{author}{\bibinfo{person}{Frank~J Massey~Jr}.} \bibinfo{year}{1951}\natexlab{}.
\newblock \showarticletitle{The Kolmogorov-Smirnov test for goodness of fit}.
\newblock \bibinfo{journal}{\emph{Journal of the American statistical Association}} \bibinfo{volume}{46}, \bibinfo{number}{253} (\bibinfo{year}{1951}), \bibinfo{pages}{68--78}.
\newblock


\bibitem[Ong and Khong(2025)]%
        {ong2025spectrum}
\bibfield{author}{\bibinfo{person}{Rongqing~Kenneth Ong} {and} \bibinfo{person}{Andy~WH Khong}.} \bibinfo{year}{2025}\natexlab{}.
\newblock \showarticletitle{Spectrum-based Modality Representation Fusion Graph Convolutional Network for Multimodal Recommendation}. In \bibinfo{booktitle}{\emph{Proceedings of the Eighteenth ACM International Conference on Web Search and Data Mining}}. \bibinfo{pages}{773--781}.
\newblock


\bibitem[Peng et~al\mbox{.}(2024)]%
        {peng2024reconciling}
\bibfield{author}{\bibinfo{person}{Kenny Peng}, \bibinfo{person}{Manish Raghavan}, \bibinfo{person}{Emma Pierson}, \bibinfo{person}{Jon Kleinberg}, {and} \bibinfo{person}{Nikhil Garg}.} \bibinfo{year}{2024}\natexlab{}.
\newblock \showarticletitle{Reconciling the accuracy-diversity trade-off in recommendations}. In \bibinfo{booktitle}{\emph{Proceedings of the ACM Web Conference 2024}}. \bibinfo{pages}{1318--1329}.
\newblock


\bibitem[P{\'e}rez-Marcos et~al\mbox{.}(2020)]%
        {perez2020hybrid}
\bibfield{author}{\bibinfo{person}{Javier P{\'e}rez-Marcos}, \bibinfo{person}{Luc{\'\i}a Mart{\'\i}n-G{\'o}mez}, \bibinfo{person}{Diego~M Jim{\'e}nez-Bravo}, \bibinfo{person}{Vivian~F L{\'o}pez}, {and} \bibinfo{person}{Mar{\'\i}a~N Moreno-Garc{\'\i}a}.} \bibinfo{year}{2020}\natexlab{}.
\newblock \showarticletitle{Hybrid system for video game recommendation based on implicit ratings and social networks}.
\newblock \bibinfo{journal}{\emph{Journal of Ambient Intelligence and Humanized Computing}}  \bibinfo{volume}{11} (\bibinfo{year}{2020}), \bibinfo{pages}{4525--4535}.
\newblock


\bibitem[Peska and Dokoupil(2022)]%
        {peska2022towards}
\bibfield{author}{\bibinfo{person}{Ladislav Peska} {and} \bibinfo{person}{Patrik Dokoupil}.} \bibinfo{year}{2022}\natexlab{}.
\newblock \showarticletitle{Towards results-level proportionality for multi-objective recommender systems}. In \bibinfo{booktitle}{\emph{Proceedings of the 45th International ACM SIGIR Conference on Research and Development in Information Retrieval}}. \bibinfo{pages}{1963--1968}.
\newblock


\bibitem[Sun et~al\mbox{.}(2024)]%
        {sun2024self}
\bibfield{author}{\bibinfo{person}{Youchen Sun}, \bibinfo{person}{Zhu Sun}, \bibinfo{person}{Yingpeng Du}, \bibinfo{person}{Jie Zhang}, {and} \bibinfo{person}{Yew~Soon Ong}.} \bibinfo{year}{2024}\natexlab{}.
\newblock \showarticletitle{Self-Supervised Denoising through Independent Cascade Graph Augmentation for Robust Social Recommendation}. In \bibinfo{booktitle}{\emph{Proceedings of the 30th ACM SIGKDD Conference on Knowledge Discovery and Data Mining}}. \bibinfo{pages}{2806--2817}.
\newblock


\bibitem[Tao et~al\mbox{.}(2022)]%
        {tao2022self}
\bibfield{author}{\bibinfo{person}{Zhulin Tao}, \bibinfo{person}{Xiaohao Liu}, \bibinfo{person}{Yewei Xia}, \bibinfo{person}{Xiang Wang}, \bibinfo{person}{Lifang Yang}, \bibinfo{person}{Xianglin Huang}, {and} \bibinfo{person}{Tat-Seng Chua}.} \bibinfo{year}{2022}\natexlab{}.
\newblock \showarticletitle{Self-supervised learning for multimedia recommendation}.
\newblock \bibinfo{journal}{\emph{IEEE Transactions on Multimedia}}  \bibinfo{volume}{25} (\bibinfo{year}{2022}), \bibinfo{pages}{5107--5116}.
\newblock


\bibitem[Tian et~al\mbox{.}(2022a)]%
        {tian2022learning}
\bibfield{author}{\bibinfo{person}{Changxin Tian}, \bibinfo{person}{Yuexiang Xie}, \bibinfo{person}{Yaliang Li}, \bibinfo{person}{Nan Yang}, {and} \bibinfo{person}{Wayne~Xin Zhao}.} \bibinfo{year}{2022}\natexlab{a}.
\newblock \showarticletitle{Learning to denoise unreliable interactions for graph collaborative filtering}. In \bibinfo{booktitle}{\emph{Proceedings of the 45th international ACM SIGIR conference on research and development in information retrieval}}. \bibinfo{pages}{122--132}.
\newblock


\bibitem[Tian et~al\mbox{.}(2022b)]%
        {tian2022reciperec}
\bibfield{author}{\bibinfo{person}{Yijun Tian}, \bibinfo{person}{Chuxu Zhang}, \bibinfo{person}{Zhichun Guo}, \bibinfo{person}{Chao Huang}, \bibinfo{person}{Ronald Metoyer}, {and} \bibinfo{person}{Nitesh~V Chawla}.} \bibinfo{year}{2022}\natexlab{b}.
\newblock \showarticletitle{RecipeRec: A heterogeneous graph learning model for recipe recommendation}.
\newblock \bibinfo{journal}{\emph{arXiv preprint arXiv:2205.14005}} (\bibinfo{year}{2022}).
\newblock


\bibitem[Wei et~al\mbox{.}(2023)]%
        {wei2023multi}
\bibfield{author}{\bibinfo{person}{Wei Wei}, \bibinfo{person}{Lianghao Xia}, {and} \bibinfo{person}{Chao Huang}.} \bibinfo{year}{2023}\natexlab{}.
\newblock \showarticletitle{Multi-relational contrastive learning for recommendation}. In \bibinfo{booktitle}{\emph{Proceedings of the 17th ACM conference on recommender systems}}. \bibinfo{pages}{338--349}.
\newblock


\bibitem[Wen et~al\mbox{.}(2023)]%
        {wen2023graph}
\bibfield{author}{\bibinfo{person}{Jie Wen}, \bibinfo{person}{Gehui Xu}, \bibinfo{person}{Zhanyan Tang}, \bibinfo{person}{Wei Wang}, \bibinfo{person}{Lunke Fei}, {and} \bibinfo{person}{Yong Xu}.} \bibinfo{year}{2023}\natexlab{}.
\newblock \showarticletitle{Graph regularized and feature aware matrix factorization for robust incomplete multi-view clustering}.
\newblock \bibinfo{journal}{\emph{IEEE Transactions on Circuits and Systems for Video Technology}} \bibinfo{volume}{34}, \bibinfo{number}{5} (\bibinfo{year}{2023}), \bibinfo{pages}{3728--3741}.
\newblock


\bibitem[Williams(1991)]%
        {williams1991probability}
\bibfield{author}{\bibinfo{person}{David Williams}.} \bibinfo{year}{1991}\natexlab{}.
\newblock \bibinfo{booktitle}{\emph{Probability with martingales}}.
\newblock \bibinfo{publisher}{Cambridge university press}.
\newblock


\bibitem[Xiao and Jiang(2024)]%
        {xiao2024divnet}
\bibfield{author}{\bibinfo{person}{Shuai Xiao} {and} \bibinfo{person}{Zaifan Jiang}.} \bibinfo{year}{2024}\natexlab{}.
\newblock \showarticletitle{DivNet: Diversity-Aware Self-Correcting Sequential Recommendation Networks}. In \bibinfo{booktitle}{\emph{Proceedings of the 33rd ACM International Conference on Information and Knowledge Management}}. \bibinfo{pages}{5000--5006}.
\newblock


\bibitem[Xu et~al\mbox{.}(2024)]%
        {xu2024learning}
\bibfield{author}{\bibinfo{person}{Yuanbo Xu}, \bibinfo{person}{Fuzhen Zhuang}, \bibinfo{person}{En Wang}, \bibinfo{person}{Chaozhuo Li}, {and} \bibinfo{person}{Jie Wu}.} \bibinfo{year}{2024}\natexlab{}.
\newblock \showarticletitle{Learning without Missing-At-Random Prior Propensity-A Generative Approach for Recommender Systems}.
\newblock \bibinfo{journal}{\emph{IEEE Transactions on Knowledge and Data Engineering}} (\bibinfo{year}{2024}).
\newblock


\bibitem[Yang et~al\mbox{.}(2022a)]%
        {yang2022large}
\bibfield{author}{\bibinfo{person}{Liangwei Yang}, \bibinfo{person}{Zhiwei Liu}, \bibinfo{person}{Yu Wang}, \bibinfo{person}{Chen Wang}, \bibinfo{person}{Ziwei Fan}, {and} \bibinfo{person}{Philip~S Yu}.} \bibinfo{year}{2022}\natexlab{a}.
\newblock \showarticletitle{Large-scale personalized video game recommendation via social-aware contextualized graph neural network}. In \bibinfo{booktitle}{\emph{Proceedings of the ACM Web Conference 2022}}. \bibinfo{pages}{3376--3386}.
\newblock


\bibitem[Yang et~al\mbox{.}(2022b)]%
        {SCGRec}
\bibfield{author}{\bibinfo{person}{Liangwei Yang}, \bibinfo{person}{Zhiwei Liu}, \bibinfo{person}{Yu Wang}, \bibinfo{person}{Chen Wang}, \bibinfo{person}{Ziwei Fan}, {and} \bibinfo{person}{Philip~S. Yu}.} \bibinfo{year}{2022}\natexlab{b}.
\newblock \showarticletitle{Large-scale Personalized Video Game Recommendation via Social-aware Contextualized Graph Neural Network}. In \bibinfo{booktitle}{\emph{{WWW} '22: The {ACM} Web Conference 2022, Virtual Event, Lyon, France, April 25 - 29, 2022}}. \bibinfo{publisher}{{ACM}}, \bibinfo{pages}{3376--3386}.
\newblock
\href{https://doi.org/10.1145/3485447.3512273}{doi:\nolinkurl{10.1145/3485447.3512273}}


\bibitem[Yang et~al\mbox{.}(2023)]%
        {yang2023dgrec}
\bibfield{author}{\bibinfo{person}{Liangwei Yang}, \bibinfo{person}{Shengjie Wang}, \bibinfo{person}{Yunzhe Tao}, \bibinfo{person}{Jiankai Sun}, \bibinfo{person}{Xiaolong Liu}, \bibinfo{person}{Philip~S Yu}, {and} \bibinfo{person}{Taiqing Wang}.} \bibinfo{year}{2023}\natexlab{}.
\newblock \showarticletitle{Dgrec: Graph neural network for recommendation with diversified embedding generation}. In \bibinfo{booktitle}{\emph{Proceedings of the sixteenth ACM international conference on web search and data mining}}. \bibinfo{pages}{661--669}.
\newblock


\bibitem[Ye et~al\mbox{.}(2021)]%
        {ye2021dynamic}
\bibfield{author}{\bibinfo{person}{Rui Ye}, \bibinfo{person}{Yuqing Hou}, \bibinfo{person}{Te Lei}, \bibinfo{person}{Yunxing Zhang}, \bibinfo{person}{Qing Zhang}, \bibinfo{person}{Jiale Guo}, \bibinfo{person}{Huaiwen Wu}, {and} \bibinfo{person}{Hengliang Luo}.} \bibinfo{year}{2021}\natexlab{}.
\newblock \showarticletitle{Dynamic graph construction for improving diversity of recommendation}. In \bibinfo{booktitle}{\emph{Proceedings of the 15th ACM Conference on Recommender Systems}}. \bibinfo{pages}{651--655}.
\newblock


\bibitem[Yin et~al\mbox{.}(2024)]%
        {yin2024simple}
\bibfield{author}{\bibinfo{person}{Qing Yin}, \bibinfo{person}{Hui Fang}, \bibinfo{person}{Zhu Sun}, {and} \bibinfo{person}{Yew-Soon Ong}.} \bibinfo{year}{2024}\natexlab{}.
\newblock \showarticletitle{A Simple Yet Effective Approach for Diversified Session-Based Recommendation}.
\newblock \bibinfo{journal}{\emph{arXiv preprint arXiv:2404.00261}} (\bibinfo{year}{2024}).
\newblock


\bibitem[Yu et~al\mbox{.}(2023)]%
        {yu2023multi}
\bibfield{author}{\bibinfo{person}{Penghang Yu}, \bibinfo{person}{Zhiyi Tan}, \bibinfo{person}{Guanming Lu}, {and} \bibinfo{person}{Bing-Kun Bao}.} \bibinfo{year}{2023}\natexlab{}.
\newblock \showarticletitle{Multi-view graph convolutional network for multimedia recommendation}. In \bibinfo{booktitle}{\emph{Proceedings of the 31st ACM international conference on multimedia}}. \bibinfo{pages}{6576--6585}.
\newblock


\bibitem[Zhao et~al\mbox{.}(2024a)]%
        {zhao2024denoising}
\bibfield{author}{\bibinfo{person}{Jujia Zhao}, \bibinfo{person}{Wang Wenjie}, \bibinfo{person}{Yiyan Xu}, \bibinfo{person}{Teng Sun}, \bibinfo{person}{Fuli Feng}, {and} \bibinfo{person}{Tat-Seng Chua}.} \bibinfo{year}{2024}\natexlab{a}.
\newblock \showarticletitle{Denoising diffusion recommender model}. In \bibinfo{booktitle}{\emph{Proceedings of the 47th International ACM SIGIR Conference on Research and Development in Information Retrieval}}. \bibinfo{pages}{1370--1379}.
\newblock


\bibitem[Zhao et~al\mbox{.}(2024b)]%
        {zhao2024can}
\bibfield{author}{\bibinfo{person}{Yuying Zhao}, \bibinfo{person}{Minghua Xu}, \bibinfo{person}{Huiyuan Chen}, \bibinfo{person}{Yuzhong Chen}, \bibinfo{person}{Yiwei Cai}, \bibinfo{person}{Rashidul Islam}, \bibinfo{person}{Yu Wang}, {and} \bibinfo{person}{Tyler Derr}.} \bibinfo{year}{2024}\natexlab{b}.
\newblock \showarticletitle{Can one embedding fit all? a multi-interest learning paradigm towards improving user interest diversity fairness}. In \bibinfo{booktitle}{\emph{Proceedings of the ACM Web Conference 2024}}. \bibinfo{pages}{1237--1248}.
\newblock


\bibitem[Zheng et~al\mbox{.}(2021)]%
        {zheng2021dgcn}
\bibfield{author}{\bibinfo{person}{Yu Zheng}, \bibinfo{person}{Chen Gao}, \bibinfo{person}{Liang Chen}, \bibinfo{person}{Depeng Jin}, {and} \bibinfo{person}{Yong Li}.} \bibinfo{year}{2021}\natexlab{}.
\newblock \showarticletitle{DGCN: Diversified recommendation with graph convolutional networks}. In \bibinfo{booktitle}{\emph{Proceedings of the Web Conference 2021}}. \bibinfo{pages}{401--412}.
\newblock


\bibitem[Zheng et~al\mbox{.}(2024)]%
        {zheng2024diversity}
\bibfield{author}{\bibinfo{person}{Yongsen Zheng}, \bibinfo{person}{Guohua Wang}, \bibinfo{person}{Yang Liu}, {and} \bibinfo{person}{Liang Lin}.} \bibinfo{year}{2024}\natexlab{}.
\newblock \showarticletitle{Diversity Matters: User-Centric Multi-Interest Learning for Conversational Movie Recommendation}. In \bibinfo{booktitle}{\emph{Proceedings of the 32nd ACM International Conference on Multimedia}}. \bibinfo{pages}{9515--9524}.
\newblock


\bibitem[Zhou et~al\mbox{.}(2023)]%
        {zhou2023bootstrap}
\bibfield{author}{\bibinfo{person}{Xin Zhou}, \bibinfo{person}{Hongyu Zhou}, \bibinfo{person}{Yong Liu}, \bibinfo{person}{Zhiwei Zeng}, \bibinfo{person}{Chunyan Miao}, \bibinfo{person}{Pengwei Wang}, \bibinfo{person}{Yuan You}, {and} \bibinfo{person}{Feijun Jiang}.} \bibinfo{year}{2023}\natexlab{}.
\newblock \showarticletitle{Bootstrap latent representations for multi-modal recommendation}. In \bibinfo{booktitle}{\emph{Proceedings of the ACM web conference 2023}}. \bibinfo{pages}{845--854}.
\newblock


\bibitem[Ziegler et~al\mbox{.}(2005)]%
        {ziegler2005improving}
\bibfield{author}{\bibinfo{person}{Cai-Nicolas Ziegler}, \bibinfo{person}{Sean~M McNee}, \bibinfo{person}{Joseph~A Konstan}, {and} \bibinfo{person}{Georg Lausen}.} \bibinfo{year}{2005}\natexlab{}.
\newblock \showarticletitle{Improving recommendation lists through topic diversification}. In \bibinfo{booktitle}{\emph{Proceedings of the 14th international conference on World Wide Web}}. \bibinfo{pages}{22--32}.
\newblock


\end{thebibliography}
\end{document}